\documentclass[
]{revtex4-1}

\draft 
\usepackage{amsmath,amssymb}
\usepackage{mathtools}
\usepackage{bm}
\usepackage{bbold}
\usepackage{fullpage}

\usepackage{tikz}
\usepackage{pgfplots}
\usepackage{color}
\usepackage{scalefnt}
\usepackage{graphicx}
\usepackage{subfigure}

\newcommand{\protein}[1]{x_{#1}}
\newcommand{\prot}[1]{\delta x_{#1}}
\newcommand{\protnondim}[1]{\delta \tilde{x}_{#1}}
\newcommand{\y}[1]{y_{#1}}
\newcommand{\km}[3]{k^-_{{#1},#2#3}}
\newcommand{\kp}[3]{k^+_{#1#2,{#3}}}

\newcommand{\T}{\rm{T}}

\newcommand{\lblock}[2]{L^{\rm{{#1}},\rm{{#2}}}}
\newcommand{\lblockb}[2]{\bm{L}^{\rm{{#1}},\rm{{#2}}}}

\newcommand{\order}{\mathcal{O}}

\newcommand{\bdx}{\bm{\delta x}}

\newcommand{\dt}{\Delta t}

\newcommand{\substrate}{u}
\newcommand{\product}{p}
\newcommand{\enzyme}{e}
\newcommand{\enzymep}{e'}
\newcommand{\complex}{c}
\newcommand{\ksp}{k^+_{ue,{c}}}
\newcommand{\ksm}{k^-_{{c},ue}}
\newcommand{\kpp}{k^+_{pe,{c}}}
\newcommand{\kpm}{k^-_{{c},pe}}
\newcommand{\etot}{x_e^{\rm{tot}}}
\newcommand{\vmax}{V_{\rm{max}}}
\newcommand{\Km}{K_{\rm{m}}}
\newcommand{\vmaxs}{V_{\rm{u}}}
\newcommand{\vmaxp}{V_{\rm{p}}}
\newcommand{\Kms}{K_{\rm{u}}}
\newcommand{\Kmp}{K_{\rm{p}}}
\newcommand{\kspt}{\bar{k}^+_{ue,{c}}}
\newcommand{\ksmt}{\bar{k}^-_{{c},ue}}
\newcommand{\kpmt}{\bar{k}^-_{{c},pe}}
\newcommand{\etott}{\bar{x}_e^{\rm{tot}}}
\newcommand{\fast}{\gamma}

\newcommand{\kmtil}[3]{\bar{k}^-_{#1,#2#3}}
\newcommand{\kptil}[3]{\bar{k}^+_{#1#2,#3}}
\newcommand{\ytil}[1]{\bar{y}_{#1}}
\newcommand{\Le}{\bm{L}_{\setminus\rm{\enzyme}}}
\newcommand{\dLe}{\Delta\Le}
\newcommand{\Leff}{\bm{L}_{\rm eff}}

\newcommand{\etil}{\tilde{e}}
\newcommand{\etilp}{\tilde{e}'}
\newcommand{\stil}{\tilde{s}}
\newcommand{\btil}{\tilde{b}}

\begin{document}


\title{Michaelis-Menten dynamics in protein subnetworks} 



\author{Katy J. Rubin}
\author{Peter Sollich}
\email[]{peter.sollich@kcl.ac.uk}
\affiliation{Department of Mathematics, King's College London, Strand, London, WC2R 2LS, UK}


\date{\today}

\begin{abstract}
To understand the behaviour of complex systems it is often necessary to use models that describe the dynamics of subnetworks. It has previously been established using projection methods that such subnetwork dynamics generically involves memory of the past, and that the memory functions can be calculated explicitly for biochemical reaction networks made up of unary and binary reactions.
However, many established network models involve also Michaelis-Menten kinetics, to describe e.g.\ enzymatic reactions. We show that the projection approach to subnetwork dynamics can be extended to such networks, thus significantly broadening its range of applicability. To derive the extension we construct a larger network that represents enzymes and enzyme complexes explicitly, obtain the projected equations, and finally take the limit of fast enzyme reactions that gives back Michaelis-Menten kinetics. The crucial point is that this limit can be taken in closed form. The outcome is a simple procedure that allows one to obtain a description of subnetwork dynamics, including memory functions, starting directly from any given network of unary, binary and Michaelis-Menten reactions. Numerical tests show that this closed form enzyme elimination gives a much more accurate description of the subnetwork dynamics than the simpler method that represents enzymes explicitly, and is also more efficient computationally.
\end{abstract}


\maketitle 


\section{Introduction}

Biological networks are often complex and it is frequently necessary to focus on subnetworks of a larger system, e.g.\ to enable a better understanding of the network properties \cite{Bhalla2003,Ackermann2012,Conradi2007}. Such subnetworks may be of interest because they carry out important biological functions or because they capture parts of the system where there is less uncertainty in the network structure or dynamical parameters. The choice of subnetwork may also be dictated by experimental data only being available for a limited number of molecular species.

There are many different methods of model reduction that have been used to simplify a large model down to one for a subnetwork \cite{Okino1998,Radulescu2012}. We have previously used projection techniques to find a systematic description of the dynamics of a subnetwork embedded in a ``bulk'' network \cite{Rubin2014d}, identifying the occurrence of memory terms as one of the key features. The resulting methods for calculating memory functions were, however, restricted to networks involving only unary and binary reactions with concentration-independent reaction rates.
While this class of networks is large, it excludes networks with reactions following Michaelis-Menten kinetics, which is commonly used to represent e.g.\ enzyme reactions. Our aim in this paper is to remove this restriction, by providing an explicit method for constructing the description of the dynamics of any subnetwork within a network consisting of unary, binary and Michaelis-Menten reactions.

An approximate way of achieving the above aim is to extend a network with Michaelis-Menten reactions to one that explicitly involves enzymes and enzyme complexes. The Michaelis-Menten terms in the kinetics are thus ``unfolded'' into unary and binary reactions so that the original projection approach \cite{Rubin2014d} can be applied. However, this is inconvenient both conceptually -- we need to include extra species not present in the original system of reaction equations -- and numerically, because the fast rates of enzyme reactions typically create a ``stiff'' system of differential equations that has to be integrated using small timesteps. The approach nevertheless provides the inspiration for the method we adopt here: we follow the above route {\em analytically}, taking the limit of fast enzyme reaction rates in such a way as to retrieve the original Michaelis-Menten kinetics exactly. The main achievement of our analysis is to show that the limit can be taken in closed form. This leads to a procedure for constructing the subnetwork dynamics directly from the original reaction network, without reference to any extra species.

In Section \ref{sec:MM} we give a summary of Michaelis-Menten dynamics and the conditions under which it is retrieved as the limit of a fast formation and dissociation of an enzyme complex; in essence these conditions are that the enzyme reactions must be fast, and the enzyme concentration low. Section \ref{sec:MMproj} details our approach to obtaining the projected equations that describe subnetwork dynamics, for systems of reaction equations that include Michaelis-Menten reactions: we temporarily add enzymes to represent these reactions, derive the projected equations, and then take the limit of fast enzyme rate and low enzyme concentration in closed form.
%
%
We describe the approach separately for 
linearised (Sec.~\ref{sec:linMM}) and nonlinear (Sec.~\ref{sec:nonlinMM}) dynamics, as the nonlinear case is more complicated technically but follows the same conceptual route.
Remarkably, even though Michaelis-Menten terms are generally nonlinear, we find that in the memory terms that are characteristic of the projected equations no additional nonlinearities appear, i.e.\ the memory terms involve linear concentration fluctuations for linearised dynamics, and linear and quadratic concentration fluctuations for nonlinear dynamics. Finally in Section \ref{sec:MMcomparison} we compare predictions from the original reaction equations with Michaelis-Menten dynamics to the projected equations with either enzymes added explicitly or eliminated in closed form using the method derived in this paper.
%
%
We find that the closed form elimination is both faster to evaluate computationally, and gives a more accurate approximation to the original reaction equations.

\section{Michaelis-Menten dynamics}
\label{sec:MMdynamics}

Many biochemical reactions are catalysed by enzymes. Generally each enzyme will enable a particular reaction without being consumed. A simple model of an enzyme reaction \cite{Henri1902,Michaelis1913} is written
\begin{equation}
  \label{eq:michaelismenten}
  \substrate + \enzyme \xrightleftharpoons[\ksm]{\ksp} \complex \xrightarrow{\kpm} \enzyme +\product
\end{equation}
where $\substrate$ is a substrate (we use ``u'' not ``s'' here as we will need the letter ``s'' to denote ``subnetwork'' later), $\enzyme$ is an enzyme, $\complex$ is an enzyme-substrate complex and $\product$ is a product. The reaction rates are denoted $\ksp, \ksm, \kpm$.
These reactions describe the binding of a free enzyme with a substrate to form a substrate-enzyme complex. This complex can then dissociate into enzyme and a product. In the traditional model substrate binding is reversible but product formation is not; we will consider also the more general case below, where both reactions are reversible.

\subsection{Derivation of Michaelis-Menten equations}\
\label{sec:MM}

Let $\protein{i}$ be the concentration of species $i$. Then the set of mass action equations for system \eqref{eq:michaelismenten} is
\begin{equation}
  \label{eq:MMmassaction}
  \begin{split}
    \frac{\partial}{\partial t}\protein{\substrate} &= \ksm\protein{\complex}-\ksp\protein{\substrate}\protein{\enzyme}\\
    \frac{\partial}{\partial t}\protein{\enzyme} &= \ksm\protein{\complex}-\ksp\protein{\substrate}\protein{\enzyme}+\kpm\protein{\complex}\\
    \frac{\partial}{\partial t}\protein{\complex} &= -\ksm\protein{\complex}+\ksp\protein{\substrate}\protein{\enzyme}-\kpm\protein{\complex}\\
    \frac{\partial}{\partial t}\protein{\product} &= \kpm\protein{\complex}
  \end{split}
\end{equation}
From these equations it follows that there is a conservation law between the enzyme and complex such that
\begin{equation}
  \label{eq:MMconslaw}
  \frac{\partial}{\partial t}\protein{\enzyme} + \frac{\partial}{\partial t}\protein{\complex} = 0 \Longrightarrow \protein{\enzyme}+ \protein{\complex} = \etot
\end{equation}

The Michaelis-Menten description of the dynamics is obtained by exploiting the fact that enzyme reactions are typically fast. This allows one to reduce the system \eqref{eq:michaelismenten} to a simpler description where the enzyme and enzyme complex no longer appear explicitly.

To achieve this simplification one uses the quasi-steady state assumption \cite{Briggs1925}
. We assume all enzyme reactions are fast so that the enzyme complex is in quasi-steady state at any time. We use the qualifier ``quasi'' because this steady state depends on the concentrations of both substrate and product, which themselves generally vary in time. 
For the reaction flux $v = \kpm\protein{\complex}$ one then finds
\begin{equation}
  \label{eq:MMreactionrate}
  v = \frac{\kpm\etot\protein{\substrate}}{(\ksm+\kpm)/\ksp + \protein{\substrate}} = \frac{\vmax\protein{\substrate}}{\Km+\protein{\substrate}}
\end{equation}
where
\begin{equation}
  \label{eq:vmax}
  \vmax = \kpm\etot
\end{equation}
is the maximum reaction flux that can be achieved and
\begin{equation}
  \label{eq:Km}
  \Km = \frac{\ksm+\kpm}{\ksp}
\end{equation}
is the Michaelis constant. 
The simplified description of the original reaction system can now be written in terms of the reaction flux $v$ in \eqref{eq:MMreactionrate}, as
\begin{equation}
  \label{eq:MMreactioneqns}
  \begin{split}
    -\frac{\partial}{\partial t}\protein{\substrate}&=\frac{\vmax\protein{\substrate}}{\Km+\protein{\substrate}}
=
    \frac{\partial}{\partial t}\protein{\product}
  \end{split}
\end{equation}
In the limit where the complex dissociates into enzyme and substrate at a much higher rate than for enzyme and product, one can achieve a further simplification known as the ``rapid equilibrium assumption'' \cite{Michaelis1913}. However, this is just a limiting case of the quasi-steady state assumption where $\kpm$ is neglected against $\ksm$ in determining the Michaelis constant $\Km$.




\subsection{Reversible Michaelis-Menten dynamics}
\label{sec:MMreversible}

We discuss briefly how the above analysis is modified when there is a back reaction from the enzyme and product to the complex \cite{Haldane1930,Sauro2013}. In such cases one has the more general reaction scheme
\begin{equation}
  \label{eq:MMreversible}
  \substrate + \enzyme \xrightleftharpoons[\ksm]{\ksp} \complex \xrightleftharpoons[\kpp]{\kpm} \enzyme +\product  
\end{equation}
with the additional rate constant $\kpp$. 
The equation for the enzyme complex concentration now has an extra contribution $\kpp\protein{\enzyme}\protein{\product}$, while the equations for enzyme and product contain the same additional term with a negative sign. Using the quasi-steady state assumption one then obtains a reaction flux of the form
\begin{equation}
  \label{eq:MMreactionratereverse}
  v = \frac{\vmaxs\protein{\substrate}/\Kms - \vmaxp\protein{\product}/\Kmp}{1 + \protein{\substrate}/\Kms + \protein{\product}/\Kmp}
\end{equation}
where $\vmaxs$ and $\vmaxp$ are the maximum reaction rates for the forward and reverse reactions respectively and are given by
\begin{equation}
  \label{eq:vmaxdefns}
     \vmaxs = \kpm\etot,\quad
    \vmaxp = \ksm\etot
 \end{equation}
Similarly $\Kms$ and $\Kmp$ are the Michaelis constants for the forward and reverse reactions and are written in terms of the mass action rate constants as
\begin{equation}
  \label{eq:kmdefns}
    \Kms = \frac{\ksm + \kpm}{\ksp},\quad \Kmp = \frac{\ksm+\kpm}{\kpp}
\end{equation}

\subsection{Quantitative accuracy of Michaelis-Menten approximation}
\label{sec:MMtoMA}

We outlined above how the Michaelis-Menten dynamics \eqref{eq:MMreactioneqns} -- or its generalisation \eqref{eq:MMreactionratereverse} -- emerges from a quasi-steady state approximation. We now consider under what conditions on the enzyme reaction parameters these approximations become exact, so that we can later take the appropriate limit for the enzyme parameters in the construction of the projected equations. Intuitively the enzyme reactions need to be fast; from the definition of the maximal reaction rate \eqref{eq:vmax}, which in a reaction specified from the outset in Michaelis-Menten form is given, the total enzyme concentration $\etot$ then needs to be small. This makes sense as the Michaelis-Menten kinetics always gives equal and opposite reaction fluxes for substrate and product, neglecting any substrate or product captured in the enzyme complex. A more formal analysis by \citet{Segel1989}, based on a singular perturbation approach to a dimensionless form of the reaction equations, gives $\etot/(\protein{\substrate}+\Km) \ll 1$ as the condition for validity of the quasi-steady state approximation. As $\Km$ can also be regarded as fixed by the specification of a Michaelis-Menten reaction, this is consistent with the intuitive requirement of small enzyme concentration.

The discussion so far suggests that we should rewrite the enzyme reaction rate constants and total enzyme concentrations as 
\begin{equation}
  \label{eq:reactionratesfast}
  \ksp = \kspt\fast,\ \ksm = \ksmt\fast,\ \kpm = \kpmt\fast,\ \etot = \etott/\fast
\end{equation}
where $\fast$ is a dimensionless fast rate scale parameter.
%
%
The definitions of $\vmax$ \eqref{eq:vmax} and $\Km$ \eqref{eq:Km} then allow us to write the remaining parameters as
\begin{equation}
   \etott = \vmax/\kpmt,\
\kspt= \frac{\ksmt+\kpmt}{\Km}
\end{equation}
We now have three parameters, $\ksmt$, $\kpmt$ and $\fast$, that parameterise the possible enzyme kinetics underlying a given Michaelis-Menten reaction. 
From the criterion of \cite{Segel1989} the Michaelis-Menten description will then become exact for $\fast\to\infty$, irrespective of the values of $\ksmt$ and $\kpmt$ \cite{VanSlyke1914}. The same exactness statement applies to the more general case where there is also a back reaction from product and enzyme to form an enzyme complex \cite{Li2013, Kollar2015}. 

\section{Enzyme reactions in the projected equations}
\label{sec:MMproj}

The projection method as applied to protein interaction networks \cite{Rubin2014d} works with mass action kinetics for unary and binary reactions. Therefore if we
are given an interaction network that includes Michaelis-Menten reactions then {\em a priori} we need to represent these explicitly in such a form, to allow us to compute the projected equations. As explained in the introduction, this is a disadvantage both conceptually and computationally.
Our aim here is to implement this approach {\em analytically} instead: we add enzyme species to the network to get mass action kinetics and take the large enzyme rate and low enzyme concentration limit in which the mass action dynamics becomes exactly identical to Michaelis-Menten dynamics. The challenge is to understand what happens in this limit to the projected equations.

The aim of the projection method generally is to provide a description of the dynamics of a protein interaction subnetwork embedded in a bulk network. The Zwanzig-Mori projection method \cite{Mori1965} can be used to obtain such a description, specifically equations for the time evolution of the protein concentrations in a chosen subnetwork. Full details of the projection method applied to protein interaction networks are given in \cite{Rubin2014d}. We summarise below the features necessary for the analysis of enzyme dynamics.

The natural specification of the state of a given network is the vector of concentrations for all molecular species in the network, which we call $\bm{x}$. In the steady state $\bm{x}$ will fluctuate around some mean $\bm{y}$ and it will be useful to switch variables to $\bdx=\bm{x}-\bm{y}$, which has zero mean in steady state. The protein concentrations (concentration deviations from the mean, more precisely) $\bdx$ evolve in time according to a Fokker-Planck equation, where the stochasticity arises from copy number fluctuations. The variance of these fluctuations scales as the inverse reaction (e.g.\ cell) volume $\epsilon=1/V$. While a nonzero $\epsilon$ is needed initially to apply the Zwanzig-Mori formalism, we consider the limit of small $\epsilon$ throughout.

The time evolution of any observable $a(\bdx,t)$ is given by $(\partial/\partial t)a = \mathcal{L}a$, where $\mathcal{L}$ is the adjoint Fokker-Planck operator whose drift term encodes the mass action kinetics. We showed in \cite{Rubin2014d} that if we focus on a set of observables $\bm{z}$ containing $\bdx$ and all products like $\prot{1}^2$ and $\prot{1}\prot{2}$, then the operator $\mathcal{L}$ can be written in {\em matrix form} $\bm{L}$ such that 
\begin{equation}
  \label{eq:Lmatrixeqn}
  \frac{\partial}{\partial t}z_{\alpha}=\sum_\beta z_{\beta}L_{\beta\alpha} + \prot{}^3+\order(\epsilon)
\end{equation}
where $\prot{}^3$ indicates terms of third or higher order in $\bdx$ while the $\order(\epsilon)$ terms vanish in the low noise limit.

If we now let $\{a_\alpha(\bdx)\}$ be a set of observables from the subnetwork, the projected equations have the form
\begin{equation}
  \label{eq:projectedeqns}
  \frac{\partial}{\partial t}a_\alpha(t) = \sum_\beta a_\beta(t)\Omega_{\beta\alpha} + \int_0^tdt'\,\sum_\beta a_\beta(t')M_{\beta\alpha}(t-t') + r_\alpha(t)
\end{equation}
We choose specifically as subnetwork observables all the subnetwork concentrations and their products, i.e.\ the entries of $\bm{z}$ only involving the subnetwork. We denote these collectively by S, and the remaining species -- all bulk concentrations (b) and concentration products involving the bulk (sb and bb) -- by the letter B. 

With the observables ordered appropriately, we can then decompose the matrix $\bm{L}$ formed from the entries $L_{\beta\alpha}$ into subnetwork and bulk blocks: 
\begin{equation}
\bm{L}=
  \label{eq:Lmatrix}
  \left(
\begin{array}{cc}
   \lblockb{S}{S}&   \lblockb{S}{B}\\
   \lblockb{B}{S}&\lblockb{B}{B}
 \end{array}
\right)
\end{equation}
The four blocks here can be broken down further according to the specific type of subnetwork and bulk observable, for example
\begin{equation}
\lblockb{B}{S} = 
 \left(\begin{array}{cc}
   \lblockb{b}{s}&0\\
   \lblockb{sb}{s}&\lblockb{sb}{ss}\\
   \lblockb{bb}{s}&\lblockb{bb}{ss}\\
 \end{array}\right)
\end{equation}
Here ``s'' denotes linear subnetwork observables, ``ss'' product observables from the subnetwork, ``sb'' cross-product observables between subnetwork and bulk, and so on. In this way $\lblockb{b}{s}$ contains the linear coefficients of bulk concentrations in the equations of motion for subnetwork concentrations, while the coefficients of subnetwork-bulk products in these equations are in the block $\lblockb{sb}{s}$.

In terms of the blocks of the $\bm{L}$-matrix defined as above, the ``rate matrix'' $\bm{\Omega}$ in \eqref{eq:projectedeqns} is simply \cite{Rubin2014d}
\begin{equation}
  \label{eq:omega}
 \bm{\Omega}=\lblockb{S}{S} = 
 \left(\begin{array}{cc}
   \lblockb{s}{s}&0\\
   \lblockb{ss}{s}&\lblockb{ss}{ss}
 \end{array}\right)
\end{equation}
It contains terms from the subnetwork dynamics that are local in time. The memory function (matrix) can be written as \cite{Rubin2014d}
\begin{equation}
  \label{eq:mem}
  \bm{M}(\dt) = \lblockb{S}{B}e^{\lblockb{B}{B}\dt}\lblockb{B}{S}
\end{equation}
where $\dt=t-t'$. The entries of this matrix, which appear in \eqref{eq:projectedeqns}, are the memory functions 
$M_{\beta\alpha}(\dt)$: they determine how strongly the past values of
the observable $a_\beta$ affect the present rate of change of
$a_\alpha$. For later use we note here the Laplace transform version
of the memory function, which is 
\begin{equation}
  \label{eq:memLT}
  \hat{\bm{M}}(z) = \lblockb{S}{B}\left(z-\lblockb{B}{B}\right)^{-1}\lblockb{B}{S}
\end{equation}
For brevity we will use the term ``memory function'' also for $\hat{M}(z)$ when it is clear from the context that the Laplace transform is meant.

\

One important property of the memory functions is their boundary structure: if we define a boundary species as a subnetwork species that is involved in a reaction with a bulk species, then among the projected equations for subnetwork concentrations only those for boundary species contain memory terms. We also note that linear memory functions are only nonzero for boundary species influencing other boundary species. Similarly, only concentration products involving at least one boundary species appear in nonlinear memory terms \cite{Rubin2014d}. 

The final term in \eqref{eq:projectedeqns}, $r(t)$, is what is known as the random force. It accounts for the fact that because of the interaction between subnetwork and bulk, the time evolution of the subnetwork observables cannot be closed. When using the projected equations to make predictions in numerical examples, we will drop the random force as there is no simple way of calculating it. The otherwise exact projected equations thus become an approximation, but one that in previous work~\cite{Rubin2014d} we showed to be remarkably accurate.


Our analysis starts from a given reaction network involving unary and binary protein reactions, and enzyme reactions described by Michaelis-Menten terms. We assume for simplicity that all enzyme reactions are reversible; the irreversible scenario can be obtained from this as the limiting case where the rate of dissociation into enzyme and product is much larger than the rate for formation of enzyme complex in the reverse direction.
The mass-action kinetics for each enzyme reaction is
\begin{equation}
  \label{eq:MMdimensional}
\begin{split}
      \frac{\partial}{\partial t}\prot{\substrate} &= 
-f_{\substrate\enzyme,c} + \ldots\\
   \frac{\partial}{\partial t}\prot{\enzyme} &=
-f_{\substrate\enzyme,c}-f_{\product\enzyme,c}\\
    \frac{\partial}{\partial t}\prot{\product} &=
    -f_{\product\enzyme,c} + \ldots
    \end{split}
\end{equation}
where the dots indicate fluxes from other reactions. The reaction fluxes from substrate and enzyme to complex, and from product and enzyme to complex, read respectively (compare \eqref{eq:MMmassaction})
\begin{equation}
\label{eq:MMfluxes_orig}
\begin{split}
f_{\substrate\enzyme,c} &= 
      \km{c}{\substrate}{\enzyme}(-\etot+\y{\enzyme}+\prot{\enzyme}) +
      \kp{\substrate}{\enzyme}{c}
      (\y{\substrate}+\prot{\substrate})
      (\y{\enzyme} + \prot{\enzyme})\\
f_{\product\enzyme,c} &= 
      \km{c}{\product}{\enzyme}(-\etot+\y{\enzyme}+\prot{\enzyme}) +
      \kp{\product}{\enzyme}{c}
      (\y{\product}+\prot{\product})
      (\y{\enzyme} + \prot{\enzyme})
\end{split}
\end{equation}
Here we have used the enzyme conservation law \eqref{eq:MMconslaw} to eliminate the complex concentration via $\protein{\complex}=\etot-\protein{\enzyme}=
\etot-\y{\enzyme}-\prot{\enzyme}$. The enzyme steady state concentration $\y{\enzyme}$ can be found by requiring that in the steady state, where $\prot{\enzyme}=\prot{\substrate}=\prot{\product}=0$, the two fluxes must sum to zero to ensure $(\partial/\partial t)\prot{\enzyme}=0$. This gives
\begin{equation}
\label{eq:ye}
\y{\enzyme} = \etot\,
\frac{\km{c}{\substrate}{\enzyme}
      + \km{c}{\product}{\enzyme}}
{\km{c}{\substrate}{\enzyme}
      + \km{c}{\product}{\enzyme}
      + \kp{\substrate}{\enzyme}{c}
\y{\substrate}
      + \kp{\product}{\enzyme}{c}
      \y{\product}}
\end{equation}

We now write the enzymatic reaction rate constants in terms of a fast rate
scale $\fast$ as in \eqref{eq:reactionratesfast}, and similarly the steady
state enzyme concentration, which
from \eqref{eq:ye} must scale as the inverse
of $\fast$, i.e.\ $\y{\enzyme}=\ytil{\enzyme}/\fast$, 
like the total enzyme concentration $\etot$. This gives 
\begin{equation}
\label{eq:MMfluxes}
\begin{split}
f_{\substrate\enzyme,c} &= 
      \kmtil{c}{\substrate}{\enzyme}\fast(-\etott/\fast+\ytil{\enzyme}/\fast+\prot{\enzyme}) +
      \kptil{\substrate}{\enzyme}{c}\fast
      (\y{\substrate}+\prot{\substrate})
      (\ytil{\enzyme}/\fast + \prot{\enzyme})\\
f_{\product\enzyme,c} &= 
      \kmtil{c}{\product}{\enzyme}\fast(-\etott/\fast+\ytil{\enzyme}/\fast+\prot{\enzyme}) +
      \kptil{\product}{\enzyme}{c}\fast
      (\y{\product}+\prot{\product})
      (\ytil{\enzyme}/\fast + \prot{\enzyme})
\end{split}
\end{equation}
The scaled rate constants and steady state enzyme concentration are related to the Michaelis-Menten parameters as explained after \eqref{eq:MMreversible}, i.e.\ 
\begin{equation}
\label{eq:MMrev_constants_vs_rates}
\begin{split}
    \vmaxs &= \frac{\kmtil{c}{p}{e}\ytil{\enzyme}(\kmtil{c}{\substrate}{\enzyme}
      + \kmtil{c}{\product}{\enzyme}
      + \kptil{\substrate}{\enzyme}{c}
\y{\substrate}
      + \kptil{\product}{\enzyme}{c}
      \y{\product})}{\kmtil{c}{\substrate}{\enzyme}
      + \kmtil{c}{\product}{\enzyme}},
\quad \Kms = \frac{\kmtil{c}{\product}{\enzyme}+\kmtil{c}{\substrate}{\enzyme}}{\kptil{\substrate}{\enzyme}{c}}\\
    \vmaxp &= \frac{\kmtil{c}{u}{e}\ytil{\enzyme}(\kmtil{c}{\substrate}{\enzyme}
      + \kmtil{c}{\product}{\enzyme}
      + \kptil{\substrate}{\enzyme}{c}
\y{\substrate}
      + \kptil{\product}{\enzyme}{c}
      \y{\product})}{\kmtil{c}{\substrate}{\enzyme}
      + \kmtil{c}{\product}{\enzyme}},\quad
\Kmp = \frac{\kmtil{c}{\product}{\enzyme}+\kmtil{c}{\substrate}{\enzyme}}{\kptil{\product}{\enzyme}{c}}
\end{split}
\end{equation}
In the above representation it is not obvious which terms have to be regarded as ``fast'' in the remainder of the analysis, and which as slow. We therefore switch to dimensionless concentration variables $\protnondim{i}=\prot{i}/\y{i}$. In terms of these we have
\begin{equation}
  \label{eq:MMdimensionless}
\begin{split}
      \frac{\partial}{\partial t}\protnondim{\substrate} &= 
-f_{\substrate\enzyme,c}/\y{\substrate} + \ldots\\
   \frac{\partial}{\partial t}\protnondim{\enzyme} &=
-\fast(f_{\substrate\enzyme,c}+f_{\product\enzyme,c})/\ytil{\enzyme} \\
    \frac{\partial}{\partial t}\protnondim{\product} &=
    -f_{\product\enzyme,c}/\y{\product} + \ldots
    \end{split}
\end{equation}
with enzymatic fluxes
\begin{equation}
\label{eq:MMfluxes_dimensionless}
\begin{split}
f_{\substrate\enzyme,c} &= 
      \kmtil{c}{\substrate}{\enzyme}\ytil{\enzyme} (-\etott/\ytil{\enzyme}+1+\protnondim{\enzyme}) +
      \kptil{\substrate}{\enzyme}{c}\y{\substrate}\ytil{\enzyme}
      (1+\protnondim{\substrate})
      (1 + \protnondim{\enzyme})\\
f_{\product\enzyme,c} &= 
      \kmtil{c}{\product}{\enzyme}\ytil{\enzyme}(-\etott/\ytil{\enzyme}+1+\protnondim{\enzyme}) +
      \kptil{\product}{\enzyme}{c}\y{\product}\ytil{\enzyme}
      (1+\protnondim{\product})
      (1 + \protnondim{\enzyme})
\end{split}
\end{equation}
Here one sees clearly that the enzyme evolution equation contains only fast terms that scale with $\fast$, while the equations for substrate and product only contain slow terms. We will therefore use dimensionless concentrations throughout, and to lighten the notation we will in the following drop the tildes, as well as the bars indicating rescaling with $\fast$. Note also that as at steady state ($\bdx=0$) the total flux into or out of any molecular species must be zero, we can drop the constant pieces from $f_{\substrate\enzyme,c}$ and
$f_{\product\enzyme,c}$: they
have to cancel against each other in the equation for $\prot{\enzyme}$, and against the other steady state fluxes in the
equations for $\prot{\substrate}$ and $\prot{\product}$. 

\begin{figure}[!ht]
  \centering
\mbox{
\subfigure[\ subnetwork]{\includegraphics[scale=0.8]{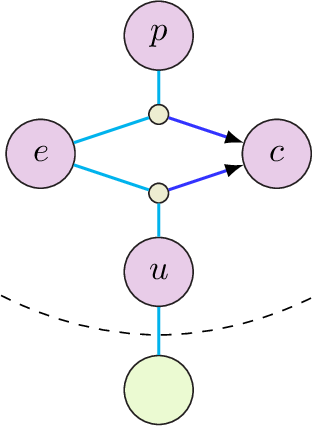}\label{fig:enzymesub}}
\quad
\subfigure[\ bulk enzymes away from the boundary]{\includegraphics[scale=0.8]{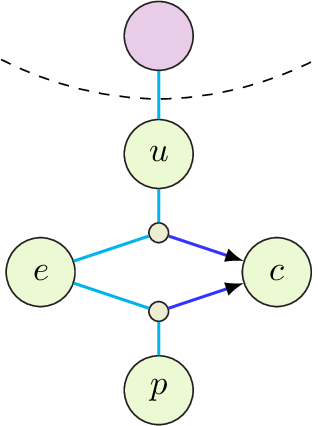}\label{fig:enzymebulk}}
\quad
\subfigure[\ bulk enzymes at the boundary]{\includegraphics[scale=0.8]{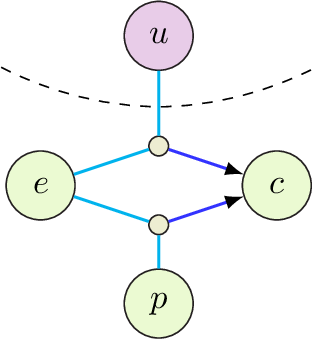}\label{fig:enzymebound}}
}
  \caption[Example enzyme networks]{The three possible locations of enzyme reactions with respect to subnetwork and bulk. (a) Substrate and product, and hence entire reaction, in the subnetwork;  (b) substrate and product, and hence entire reaction, in the bulk; (c) substrate in the subnetwork and product in the bulk, in which case we assign also the enzyme and enzyme complex to the bulk.}
  \label{fig:3enzymecases}
\end{figure}

Our task now is to take a mass-action (unary and binary) reaction system where every
enzyme reaction is represented as in
\eqref{eq:MMfluxes_dimensionless}, and to find closed form expressions
for the rate matrix and memory functions in the limit
$\fast\to\infty$ where we know that this mass-action description
becomes identical to Michaelis-Menten dynamics. The effect of the
enzyme reactions depends on where they are located relative to subnetwork
and bulk, with three potentially distinct cases as shown in Fig.~\ref{fig:3enzymecases}.
If both substrate and product are in the subnetwork, also the enzyme will be located there and, as is clear from Fig.~\ref{fig:3enzymecases}a, away from the boundary of the subnetwork. We will therefore find ``fast'' equations of motion for such enzymes without any memory terms. These enzymes can therefore be kept explicitly in a first stage of our analysis, and eliminated in a second stage following the standard logic that leads to the Michaelis-Menten description.

If both substrate and product are within the bulk then the entire enzyme reaction takes place there (Fig.~\ref{fig:3enzymecases}b), contributing fast reaction rate constants to $\lblockb{B}{B}$. Accordingly we will find that such reactions only give contributions to the memory functions, not the rate matrix.

The third case is the one where the substrate is on the boundary and the product is in the bulk (or vice versa, but for reversible enzyme reactions the two cases are mathematically equivalent). We then choose to assign also the enzyme and the enzyme complex to the bulk, and refer to this situation as a (bulk) enzyme on the boundary. Reactions involving such enzymes, whose rate constants appear in $\lblockb{S}{B},\ \lblockb{B}{B}$ and $\lblockb{B}{S}$, will contribute fast terms in the memory functions that decay on a timescale of order $1/\gamma$. In the limit $\gamma\to \infty$ these terms become local in time and so turn into contributions to the rate matrix.

\section{Linearised Dynamics}
\label{sec:linMM}

In linearised dynamics we only consider terms in the mass-action kinetics up to linear order in $\bdx$. The dimensionless scaled reaction equations for a reversible Michaelis-Menten reaction are then, from 
\eqref{eq:MMdimensionless} and \eqref{eq:MMfluxes_dimensionless},
\begin{equation}
  \label{eq:MMlin}
\begin{split}
      \frac{\partial}{\partial t}\prot{\substrate} &= -\km{c}{\substrate}{\enzyme}(\y{\enzyme}/\y{\substrate})\prot{\enzyme} - \kp{\substrate}{\enzyme}{c}\y{\enzyme}(\prot{\substrate} + \prot{\enzyme})+ \ldots\\
    \frac{\partial}{\partial t}\prot{\enzyme} &= -\fast[(\km{c}{\substrate}{\enzyme}+\km{c}{\enzyme}{\product})\prot{\enzyme} + \kp{\substrate}{\enzyme}{c}\y{\substrate}(\prot{\substrate}+\prot{\enzyme})+ \kp{\enzyme}{\product}{c}\y{\product}(\prot{\product} + \prot{\enzyme})]\\
    \frac{\partial}{\partial t}\prot{\product} &= -\km{c}{\enzyme}{\product}(\y{\enzyme}/\y{\product})\prot{\enzyme} - \kp{\enzyme}{\product}{c}\y{\enzyme}(\prot{\product} + \prot{\enzyme}) + \ldots
\end{split}
\end{equation}

Let us partition the matrix form of the adjoint Fokker-Planck matrix operator \eqref{eq:Lmatrix} so that the bulk species are split into fast and slow blocks. If e$'$ and e represent the collection of subnetwork and bulk enzymes respectively and s and b represent the other molecular species in the subnetwork and bulk, we partition as
   \begin{equation}
     \label{eq:LoperatorMMlin}
     \bm{L} =\left(
       \begin{array}{c|c}
        \lblockb{S}{S}& \lblockb{S}{B}\\
        \hline
        \lblockb{B}{S}& \lblockb{B}{B}
       \end{array}
\right)=
       \left( \begin{array}{cc|cc}
       \lblockb{s}{s}&\lblockb{s}{\enzymep}&\lblockb{s}{b}&\lblockb{s}{\enzyme}\\
       \lblockb{\enzymep}{s}&\lblockb{\enzymep}{\enzymep}&\lblockb{\enzymep}{b}&\lblockb{\enzymep}{\enzyme}\\
\hline
       \lblockb{b}{s}&\lblockb{b}{\enzymep}&\lblockb{b}{b}&\lblockb{b}{\enzyme}\\
       \lblockb{\enzyme}{s}&\lblockb{\enzyme}{\enzymep}&\lblockb{\enzyme}{b}&\lblockb{\enzyme}{\enzyme}
     \end{array}\right)
=\left(     \begin{array}{c|cc}
m &w_1&f_1\\ \hline
w_2&w_3&f_2\\
w_4&w_5&f_3\\
\end{array}\right)
    \end{equation}
where $w$ are slow terms and $f$ are fast terms; the top left block denoted $m$ contains a mixture of fast and slow terms. In writing the last equality above we have grouped s and e$'$ together; the resulting specific $3\times3$ block structure of slow and fast terms is one that we will find again in the case of the full nonlinear dynamics.
Note that because subnetwork enzymes only have interactions with subnetwork species (s and e$'$), $\lblockb{b}{\enzymep}$ and $\lblockb{\enzyme}{\enzymep}$ are zero. Similarly, because bulk proteins or enzymes do not interact with subnetwork enzymes, $\lblockb{\enzymep}{b}$ and $\lblockb{\enzymep}{\enzyme}$ vanish. This means that subnetwork enzymes do not feature at all in the calculation of the memory function \eqref{eq:memLT}, which makes intuitive sense. The vanishing of $\lblockb{b}{\enzymep}$ and $\lblockb{\enzyme}{\enzymep}$ is important also as these blocks contain rate constants for the time evolution of (subnetwork) enzymes, which by our construction scale with $\fast$: if these blocks were nonzero, it would change the character of $w_2$ and $w_4$ from slow to fast.

\begin{figure}[!ht]
  \centering
\subfigure[]{\includegraphics[scale=0.8]{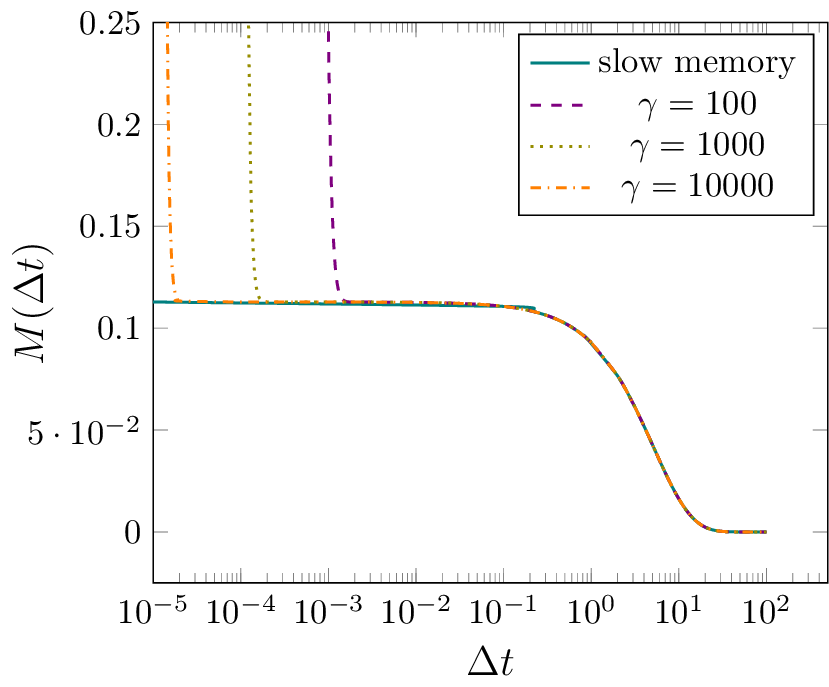}}
\quad
\subfigure[]{\includegraphics[scale=0.8]{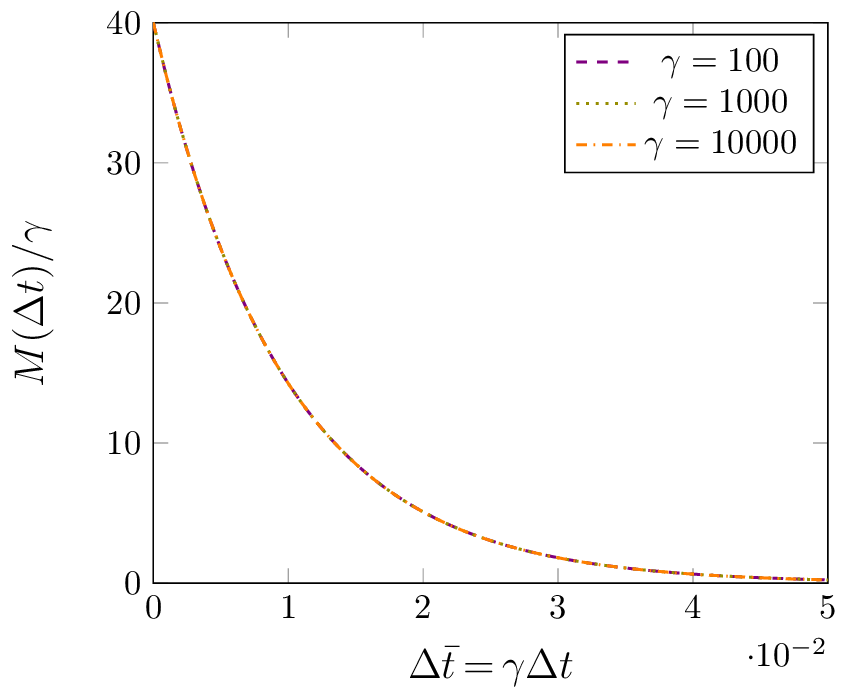}}
   \caption[Memory function with enzyme on the boundary]{Example plots of a memory function when the enzyme is on the boundary, for different values of $\fast$. (a) The memory has a slow and a fast part, with the fast part moving to shorter and shorter timescales as $\fast$ increases. 
     (b) 
     Scaled plot of the fast part: if the fast memory is divided by $\fast$ we see that a scaling plot is approached for large $\fast$. This shows that the amplitude of the fast part of the memory grows proportionally to $\fast$.}
  \label{fig:fastmemory}
\end{figure}

To analyse the memory function \eqref{eq:mem} that results from \eqref{eq:LoperatorMMlin}, we note that $\lblockb{B}{B}$ has both slow and fast sub-blocks. As a result the memory function should in general have both slow contributions that decay on $O(1)$ timescales, and fast contributions that decay for time differences of $O(1/\fast)$. As the memory function appears as a weight in an integral over the past \eqref{eq:projectedeqns}, the fast contributions only matter for $\fast\to\infty$ if their amplitude is proportional to $\fast$ so that the integral over all time differences remains finite. Accounting also for subleading terms in the amplitude dependence then suggests the following decomposition of the memory function: 
\begin{equation}
  \label{eq:memt}
  M(\dt) = (\fast M_f^0(\Delta\bar{t}) + M_f^1(\Delta\bar{t}) + \ldots) + (M_w^0(\dt) + \frac{1}{\fast}M_w^1(\dt) + \ldots)
\end{equation}
where $\Delta\bar{t}=\fast \dt$. In principle an arbitrary constant can be added to e.g.\ $M_f^1(\Delta\bar{t})$ and subtracted from $M_w^0(\dt)$; we fix this constant by requiring that all fast contributions decay to zero for large $\Delta\bar{t}$. Fig.~\ref{fig:fastmemory} shows an example of the above decomposition.

The leading fast and slow contributions can now be extracted relatively simply from the Laplace transform of the memory function \eqref{eq:memLT}, which has the decomposition
\begin{equation}
  \label{eq:memz}
  \hat{M}(z) = (\hat{M}_f^0(\bar{z}) + \frac{1}{\fast}\hat{M}_f^1(\bar{z}) + \ldots) + (\hat{M}_w^0(z) + \frac{1}{\fast}\hat{M}_w^1(z)+\ldots)
\end{equation}
where $\bar{z}=z/\fast$.
The leading fast term can be obtained by taking the limit
\begin{equation}
  \label{eq:memfast}
  \lim_{\fast\to \infty}\hat{M}(z)\big|_{\bar{z}=\text{const}} = \hat{M}_f^0(\bar{z})
\end{equation}
because the subleading fast terms are down by powers of $1/\fast$, and in the slow terms $z=\fast\bar{z}\to\infty$ so that the Laplace transforms $\hat{M}_w^0(z)$ etc.\ vanish.
The leading slow part can then be found from a limit at constant $z$, namely
\begin{equation}
  \label{eq:memslow}
  \lim_{\fast\to \infty}\left[\hat{M}(z)- \hat{M}_f^0(0)\right]\Big|_{z=\text{const}} = \hat{M}_w^0(z)
\end{equation}
We could have equivalently written $\hat{M}_f^0(\bar{z})$ inside the square brackets, as $\bar{z}=z/\fast\to 0$ when $\fast\to\infty$ at fixed $z$.

Using the method above, we can now find the fast and slow pieces of the memory function derived from the adjoint Fokker-Planck (matrix) operator in equation \eqref{eq:LoperatorMMlin}. From  \eqref{eq:memLT} this memory function reads 
\begin{equation}
  \label{eq:MMmemLTorig}
  \hat{M}(z) =  \begin{pmatrix}
    w_1 & f_1
  \end{pmatrix}
  \begin{pmatrix}
    z - w_3 & - f_2\\
    - w_5 & z - f_3
  \end{pmatrix}^{-1}
\begin{pmatrix}
  w_2 \\ w_4
\end{pmatrix}
\end{equation}
and after working out the inverse using standard block inversion identities 
and simplifying we obtain 
\begin{equation}
  \label{eq:MMmemLT}
\begin{split}
  \hat{M}(z)&=  \left(w_1 -f_1(z-f_3)^{-1}(- w_5)\right)\left(z-w_3 + f_2(z-f_3)^{-1}w_5\right)^{-1}\left(w_2+f_2(z-f_3)^{-1}w_4\right)\\
&\quad + f_1(z-f_3)^{-1}w_4
\end{split}
\end{equation}
If we now write all fast blocks as $f=\fast\bar{f}$, we can see that application of 
\eqref{eq:memfast} identifies the
fast part of the memory function as
\begin{equation}
  \label{eq:memzfast}
  \hat{M}_f^0(\bar{z}) = \bar{f}_1(\bar{z}-\bar{f}_3)^{-1}w_4
\end{equation}
while \eqref{eq:memslow} gives for 
the slow part
\begin{equation}
  \label{eq:memzslow}
  \hat{M}_w^0(z) = (w_1 -\bar{f}_1(\bar{f}_3)^{-1} w_5)(z-w_3  - \bar{f}_2(\bar{f}_3)^{-1}w_5)^{-1}(w_2-\bar{f}_2(\bar{f}_3)^{-1}w_4)
\end{equation}
Here we have used the fact that in the combination $z-\fast\bar{f}_3$, the first term can be neglected when $\fast\to\infty$ at constant $z$.

The fast part of the memory decays on an ever shorter timescale $\sim 1/\fast$ as $\fast$ increases. In the limit, and when it is used inside a memory function integral, it becomes equivalent to a delta function $\delta(\dt)$ multiplied by the area under the fast piece, which is just $ \hat{M}_f^0(0) = -\bar{f}_1(\bar{f}_3)^{-1}w_4$. The rate matrix that we obtain from \eqref{eq:LoperatorMMlin} for $\fast\to\infty$ is therefore
\begin{equation}
\bm{\Omega} = m -\bar{f}_1(\bar{f}_3)^{-1}w_4
\label{eq:Omega_effective}
\end{equation}
while the memory function in the same limit is given by \eqref{eq:memzslow}. We can now compare to the rate matrix and memory function that would result from an adjoint Fokker-Planck operator like \eqref{eq:LoperatorMMlin} without the fast bulk variables e, i.e.\ 
   \begin{equation}
\label{eq:L_no_e}
\Le=\left(     \begin{array}{c|c}
m &w_1\\ \hline
w_2&w_3
\end{array}\right)
    \end{equation}
This would give $\Omega=m$ and $\hat{M}(z)=w_1(z-w_3)^{-1}w_2$. Looking at \eqref{eq:memzslow} and \eqref{eq:Omega_effective}, we conclude that the large-$\fast$ limit gives results that can equivalently be obtained by using an adjoint Fokker-Planck matrix without the fast bulk variables that is modified from $\Le$ to $\Leff=\Le+\dLe$, where
\begin{equation}
\label{eq:dL}
\dLe=-\left(     \begin{array}{c|c}
\bar{f}_1(\bar{f}_3)^{-1} w_4 & \bar{f}_1(\bar{f}_3)^{-1} w_5  \\ \hline
\bar{f}_2(\bar{f}_3)^{-1}w_4 & \bar{f}_2(\bar{f}_3)^{-1}w_5
\end{array}\right)
\end{equation}
This is the key result of our first stage of elimination, which has removed the fast bulk variables.

\subsection{Bulk enzyme elimination as quasi-steady state method}
\label{sec:simplifyelim}

Before moving on to the second stage of also eliminating the subnetwork enzymes, we pause briefly to give a simpler form of our last result. While \eqref{eq:dL} gives a closed form for the effective $L$-matrix $\Le+\dLe$ we obtain after eliminating the bulk enzymes, this form is not very intuitive. We show next that there is a much simpler statement of the result, namely that $\Leff=\Le+\dLe$ can be obtained by treating all bulk enzymes as in quasi-steady state with the other molecular species.

To see this, we reinstate in the generic $3\times 3$ block structure of \eqref{eq:LoperatorMMlin} the specific notation for the linearised dynamics considered here, i.e.\
  \begin{equation}
     \bm{L} 
=\left(     \begin{array}{c|cc}
m &w_1&f_1\\ \hline
w_2&w_3&f_2\\
w_4&w_5&f_3\\
\end{array}\right)
=\left(     \begin{array}{c|cc}
\lblockb{S}{S} & \lblockb{S}{b}& \lblockb{S}{\enzyme}\\ \hline
\lblockb{b}{S} & \lblockb{b}{b}& \lblockb{b}{\enzyme}\\
\lblockb{\enzyme}{S} & \lblockb{\enzyme}{b}& \lblockb{\enzyme}{\enzyme}
\end{array}\right)
    \end{equation}
where, as before, S collects all subnetwork variables, i.e.\ subnetwork proteins s and subnetwork enzymes e$'$. 
The dynamics of the system can then be written as
\begin{equation}
\begin{split}
\frac{\partial}{\partial t}\bdx^{\text{S}^{\T}}  &= 
\bdx^{\text{S}^{\T}} \lblockb{S}{S} + \bdx^{\text{b}^{\T}} \lblockb{b}{S} + \bdx^{\text{e}^{\T}} \lblockb{e}{S}\\
\frac{\partial}{\partial t}\bdx^{\text{b}^{\T}}  &= 
\bdx^{\text{S}^{\T}} \lblockb{S}{b} + \bdx^{\text{b}^{\T}} \lblockb{b}{b} + \bdx^{\text{e}^{\T}} \lblockb{e}{b}\\
\frac{\partial}{\partial t}\bdx^{\text{e}^{\T}}  &= 
\bdx^{\text{S}^{\T}} \lblockb{S}{e} + \bdx^{\text{b}^{\T}} \lblockb{b}{e} + \bdx^{\text{e}^{\T}} \lblockb{e}{e}
\end{split}
\end{equation}
If we now impose a quasi-steady state condition for the bulk enzymes $\bdx^\text{e}$, this gives 
\begin{equation}
 \bdx^{\text{e}^{\T}} = -\left(\bdx^{\text{S}^{\T}} \lblockb{S}{e} + \bdx^{\text{b}^{\T}} \lblockb{b}{e}\right) ({\lblockb{e}{e}})^{-1}
\end{equation}
Substituting this back into the equations of motion for the subnetwork species and the bulk proteins gives 
\begin{equation}
  \begin{split}
\frac{\partial}{\partial t}\bdx^{\text{S}^{\T}}  &= 
\bdx^{\text{S}^{\T}} \left(\lblockb{S}{S}-\lblockb{S}{e} ({\lblockb{e}{e}})^{-1}\lblockb{e}{S}\right) + \bdx^{\text{b}^{\T}}\left( \lblockb{b}{S}-\lblockb{b}{e}({\lblockb{e}{e}})^{-1}\lblockb{e}{S}\right)  \\
\frac{\partial}{\partial t}\bdx^{\text{b}^{\T}}  &= 
\bdx^{\text{S}^{\T}} \left(\lblockb{S}{b}- \lblockb{S}{e}({\lblockb{e}{e}})^{-1}\lblockb{e}{b}\right) + \bdx^{\text{b}^{\T}}\left( \lblockb{b}{b} - \lblockb{b}{e} ({\lblockb{e}{e}})^{-1}\lblockb{e}{b}\right)
  \end{split}
\end{equation}
This is exactly the dynamics that is defined by the effective $L$-matrix $\Leff=\Le+\dLe$ derived above (see \eqref{eq:L_no_e} and \eqref{eq:dL}), hence proving our claim that this matrix can be constructed by imposing a quasi-steady state condition for the bulk enzymes.

\subsection{Michaelis-Menten terms as effective unary reactions}
\label{sec:linbulk}

The above bulk enzyme elimination can be carried out in closed form.
%
Each bulk enzyme can be eliminated by setting the time derivative of its concentration to zero. Here and throughout we assume that each enzyme only catalyses one reaction. (In the matrix formulation, this implies that the block $\lblockb{e}{e}$ is diagonal.) We then find for the effective dynamical equations of any substrate and product the following form:
\begin{equation}
  \label{eq:MMlin_eff}
\begin{split}
      \frac{\partial}{\partial t}\prot{\substrate} &= -\lambda_{\substrate\product}\prot{\substrate} + \lambda_{\product\substrate}(\y{\product}/\y{\substrate})\prot{\product}
     + \ldots\\
    \frac{\partial}{\partial t}\prot{\product} &= -\lambda_{\product\substrate}\prot{\product} + \lambda_{\substrate\product}(\y{\substrate}/\y{\product})\prot{\substrate} + \ldots
\end{split}
\end{equation}
where
\begin{equation}
\begin{split}
\label{eq:unary_rates}
\lambda_{\substrate\product} &= \frac{\kp{\substrate}{\enzyme}{\complex}\y{\enzyme}\left(\km{\complex}{\product}{\enzyme}+\kp{\product}{\enzyme}{\complex}\y{\product}\right)}{\km{\complex}{\substrate}{\enzyme}+\km{\complex}{\product}{\enzyme}+ \kp{\substrate}{\enzyme}{\complex}\y{\substrate}+\kp{\product}{\enzyme}{\complex}\y{\product}}\\
\lambda_{\product\substrate} &= \frac{\kp{\product}{\enzyme}{\complex}\y{\enzyme}\left(\km{\complex}{\substrate}{\enzyme}+\kp{\substrate}{\enzyme}{\complex}\y{\substrate}\right)}{\km{\complex}{\substrate}{\enzyme}+\km{\complex}{\product}{\enzyme}+ \kp{\substrate}{\enzyme}{\complex}\y{\substrate}+\kp{\product}{\enzyme}{\complex}\y{\product}}
\end{split}
 \end{equation}
are the rates for effective unary reactions converting substrate to product and back, respectively. The factors of $(\y{\product}/\y{\substrate})$ in \eqref{eq:MMlin_eff} arise because we are using dimensionless concentration variables.

The bulk enzyme elimination thus has the simple effect of replacing all bulk Michaelis-Menten reactions by unary conversion reactions with constant rates. From \eqref{eq:MMrev_constants_vs_rates} one sees that these effective rates can be expressed directly in terms of the Michaelis-Menten parameters, as
\begin{equation}
  \label{eq:unary_ratesMM}
  \begin{split}
\lambda_{\substrate\product} &= \frac{\vmaxs/\Kms + (\vmaxs+\vmaxp)/\Kms(\y{\product}/\Kmp)}{\left(1 + \y{\substrate}/\Kms + \y{\product}/\Kmp\right)^2}\\
\lambda_{\product\substrate} &= \frac{\vmaxp/\Kmp + (\vmaxs+\vmaxp)/\Kmp(\y{\substrate}/\Kms)}{\left(1 + \y{\substrate}/\Kms + \y{\product}/\Kmp\right)^2}
  \end{split}
\end{equation}
Comparing with \eqref{eq:MMreactionratereverse} shows that the rates are obtained by linearising the Michaelis-Menten reaction flux around the steady state concentrations of substrate and product. This is the closed-form procedure for bulk enzyme elimination we were after: it only requires as input the Michaelis-Menten parameters of the original network, and its steady state.

Note that the above discussion includes enzyme reactions both entirely in the bulk, or on the boundary of subnetwork and bulk (cf.\ Fig.~\ref{fig:3enzymecases}b and c). The only difference between these two cases is that for the latter group of enzymes, the effective unary reactions we have derived are between a subnetwork boundary species and a bulk species and so will contribute to the rate matrix, while for enzyme reactions entirely in the bulk the effective reactions only affect the memory function.

\subsection{Elimination of subnetwork enzymes}
\label{sec:linsub}

So far we have described how bulk enzymes can be eliminated, replacing them by effective unary conversion reactions. This allows the rate matrix and memory functions to be calculated from an effective $L$-matrix $\Leff=\Le+\dLe$. These quantities determine the projected equations of motion for the subnetwork proteins s and the subnetwork enzymes e$'$.

The second, final stage of the elimination procedure is now to eliminate the subnetwork enzymes. Elimination of the bulk enzymes does not affect the equations of motion for the subnetwork enzymes, i.e.\ these do not acquire memory terms nor are the local-in-time terms changed. 
Looking at the general formula \eqref{eq:mem} for the memory function, read in terms of the effective $L$-matrix, then confirms that the equations for the e$'$ species do not contain memory terms. This makes sense: the subnetwork enzymes are not boundary species to start with, and this is not changed by the effective unary reactions from bulk enzymes. 

The projected equation for each subnetwork enzyme thus looks exactly as the original equation in \eqref{eq:MMdimensionless}. Because of the fast rate scale $\fast$ in this, in the limit $\fast\to \infty$ each subnetwork enzyme will be in quasi-steady state with its substrate and product. Substituting the quasi-steady state enzyme concentration into the equations for substrate and product then gives again effective unary conversion reactions, with rates as given in \eqref{eq:unary_ratesMM} for the case of bulk enzymes.

\subsection{Summary of enzyme elimination procedure for linearised dynamics}

The final procedure we have arrived at for constructing projected equations for reaction systems with Michaelis-Menten terms, within linearised dynamics, is remarkably simple: replace each Michaelis-Menten term, whether in the subnetwork, the bulk or on the boundary, by its linearisation around the steady state. This gives effective rates for unary conversion reactions among each substrate-product pair (see \eqref{eq:unary_rates}).

\section{Nonlinear Dynamics}
\label{sec:nonlinMM}

We now want to extend the above approach of eliminating enzymes from the
projected equations to the full nonlinear dynamics. Directly transplanting the results from the linearised dynamics is not possible, however: if we use the quasi-steady state assumption for the enzymes as in Section~\ref{sec:linMM}, then we get back the full Michaelis-Menten nonlinearities. As these go beyond second order in $\prot{}$, they cannot be used directly in our construction of the projected equations, which starts from reaction equations with only linear and quadratic terms as appropriate for a mass action description of unary and binary reactions.

We take as our starting point the nonlinear $L$-matrix, as shown in equation \eqref{eq:Lmatrix}, but subdivide this into smaller blocks below in order to single out contributions from enzymes. 
Focussing though for now just on the distinction between linear and quadratic observables, we have two new kinds of entries. Firstly, mixed linear-quadratic elements as contained in e.g.\ $\lblockb{ss}{s}$: these are coefficients of quadratic terms in equations of motion for the concentrations (linear observables), so can be read off directly from the mass action equations. The quadratic-quadratic elements as in $\lblockb{ss}{ss}$ are coefficients from equations of motion of concentration products. For a generic product $\prot{ij}\equiv \prot{i}\prot{j}$ they are of the form
\begin{equation}
\label{eq:product_diff}
  \frac{\partial}{\partial t}\prot{i}\prot{j} = \prot{j}\frac{\partial}{\partial t}\prot{i} + \prot{i}\frac{\partial}{\partial t}\prot{j}
\end{equation}
Because we are only considering terms up to quadratic order on the r.h.s., we need to insert only the linearised equations of motion for $(\partial \prot{i}/\partial t)$ and $(\partial \prot{j}/\partial t)$. All quadratic-quadratic elements of the $L$-matrix are therefore ``inherited'' from the linearised dynamics. In particular, the above structure of the equations of motion for quadratic observables means that the time evolution of any product containing at least one enzyme factor will contain fast terms.

For the purpose of eliminating the fast degrees of freedom, the nonlinear $L$-matrix can be split into four blocks mirroring the structure of the linear $L$-matrix in equation \eqref{eq:LoperatorMMlin}, viz.
\begin{equation}
\label{eq:LoperatorMMnonlin}
\bm{L}=\left(
       \begin{array}{c|c}
        \lblockb{S}{S}& \lblockb{S}{B}\\
        \hline
        \lblockb{B}{S}& \lblockb{B}{B}
       \end{array}
\right)=
         \left( \begin{array}{cc|cc}
       \lblockb{\stil}{\stil}&\lblockb{\stil}{\etilp}&\lblockb{\stil}{\btil}&\lblockb{\stil}{\etil}\\
       \lblockb{\etilp}{\stil}&\lblockb{\etilp}{\etilp}&\lblockb{\etilp}{\btil}&\lblockb{\etilp}{\etil}\\
\hline
       \lblockb{\btil}{\stil}&\lblockb{\btil}{\etilp}&\lblockb{\btil}{\btil}&\lblockb{\btil}{\etil}\\
       \lblockb{\etil}{\stil}&\lblockb{\etil}{\etilp}&\lblockb{\etil}{\btil}&\lblockb{\etil}{\etil}
     \end{array}\right)
\end{equation}
The blocks are defined so that \~s contains the observables s and ss which have no bulk or fast factors, \~b contains the observables b, sb and bb which have at least one bulk factor but no fast factors, \~e consists of e, se, be, be$'$, ee and ee$'$ which contain at least one bulk factor and one fast factor - where the fast and bulk factor can be identical - and \~e$'$ contains e$'$, se$'$ and e$'$e$'$ where there are no bulk factors but at least one fast factor. Therefore the subnetwork block S consists of slow and fast blocks in the form of \~s and \~e$'$ respectively and similarly the bulk block B contains slow and fast contributions from \~b and \~e, respectively.

The block structure of the $L$-matrices for the linearised and nonlinear dynamics is therefore the same; however, there are some differences. The block \~e\~e contains some slow as well as fast entries, but the slow entries can be neglected in comparison in the large $\fast$ limit. 
The \~e\~e$'$ block is not zero as in the linear case due to the fact that the equation of motion for se$'$ involves products of the form be$'$ and ee$'$; importantly for our reasoning below these terms are slow, however, because of the time evolution of s in the se$'$ products.
Similarly the \~e$'$\~e block is not zero because the equation of motion for be$'$ involves slow se$'$ contributions; also in the equation for ee$'$ there are fast contributions from se$'$. The blocks \~b\~e$'$ and \~e$'$\~b remain zero, on the other hand.

We can go back to the same 3 $\times$ 3 structure for the $L$-matrix as in the linear case, by partitioning into blocks S (\~s and \~e$'$), \~b and \~e.
%
This 3 $\times$ 3 matrix can then be split into fast and slow blocks as in equation \eqref{eq:LoperatorMMlin}:
\begin{equation}
\def\arraystretch{1.3}
\bm{L} =\left(     \begin{array}{c|cc}
\lblockb{S}{S} &\lblockb{S}{\btil}&\lblockb{S}{\etil}\\ \hline
\lblockb{\btil}{S} &\lblockb{\btil}{\btil}&\lblockb{\btil}{\etil}\\
\lblockb{\etil}{S} &\lblockb{\etil}{\btil}&\lblockb{\etil}{\etil}\\
\end{array}\right)
 =\left(     \begin{array}{c|cc}
m &w_1&f_1\\ \hline
w_2&w_3&f_2\\
w_4&w_5&f_3\\
\end{array}\right) 
\end{equation}
Using the method of Sec.~\ref{sec:linMM} we can then find the slow and fast parts of the memory expressed in terms of these blocks, using exactly the same formulae as shown for the linearised dynamics in equation \eqref{eq:MMmemLT}. As before the result can be thought of as arising from a
modified $L$-matrix where all the fast bulk species and products contained in \~e are eliminated:
   \begin{equation}
\label{eq:Leff}
\Leff=
\left(     \begin{array}{c|c}
m-\bar{f}_1(\bar{f}_3)^{-1} w_4 & w_1-\bar{f}_1(\bar{f}_3)^{-1} w_5  \\ \hline
w_2-\bar{f}_2(\bar{f}_3)^{-1}w_4 & w_3-\bar{f}_2(\bar{f}_3)^{-1}w_5
\end{array}\right)
    \end{equation}
The form of $\Leff$ can be derived by eliminating the fast observables in \~e directly, by assuming that they are in steady state with respect to the fast contributions from their equations of motion. We turn next to the task of actually carrying out this elimination, which is more involved than in the linear case.

\subsection{Elimination of fast bulk variables}

We can simplify matters somewhat by noting that in order to find the correct rate matrix and memory function with enzymes eliminated, we require the equations of motion with enzymes eliminated -- and hence the relevant columns of $\Leff$ -- for the bulk observables \~b and the \emph{linear} subnetwork observables s and e$'$. We only need the linear observables s and e$'$ because to calculate the rate matrix and memory function one never requires blocks of the form $\lblockb{\cdot}{ss}$. (The intuition is that product observables can be replaced by actual products in the low noise limit we are considering; see~\cite{Rubin2014d} for details.) A further simplification comes from the fact that the original equations of motion for \~b, s and e$'$ do not depend on ee or ee$'$ and therefore we do not need to consider these observables further: we can focus on how to eliminate the remaining fast observables se, be, be$'$ and e.

We consider first the product observables se, be and be$'$. 
The equation of motion for a generic observable of type se, i.e.\ a product of the concentration of a subnetwork species $s$ with that of an enzyme $e$, reads 
\begin{equation}
  \frac{\partial}{\partial t}\prot{se} = \frac{\partial}{\partial t}\prot{s}\prot{e} = \prot{s}\frac{\partial}{\partial t}\prot{e}+\prot{e}\frac{\partial}{\partial t}\prot{s} = \prot{s}\frac{\partial}{\partial t}\prot{e}+ \text{slow}
\end{equation}
where we have used that $(\partial/\partial t)\prot{s}$ only contains slow terms. In finding the solution of the quasi-steady state condition $(\partial/\partial t)\prot{se}=0$, these slow terms can be neglected compared to the fast terms from $(\partial /\partial t)\prot{e}$. Writing the latter in the form
\begin{equation}
  \frac{\partial}{\partial t}\prot{\enzyme} = A \prot{\substrate}+ B\prot{\product}-C\prot{\enzyme}
\end{equation}
allows us to write the leading (fast) terms in the equation for the subnetwork-enzyme product as
\begin{equation}
\frac{\partial}{\partial t}\prot{s\enzyme}=  \prot{s}\frac{\partial}{\partial t}\prot{\enzyme} = A\,\prot{s\substrate}+B\,\prot{s\product}-C\,\prot{s\enzyme}
\end{equation}
Setting this to zero shows that the quasi-steady state solution is
\begin{equation}
\label{eq:se_elimination}
  \prot{se} = (A/C)\prot{s\substrate} + (B/C)\prot{s\product}
\end{equation}
Comparing with the (linear) quasi-steady state solution for the enzyme concentration itself, which is $\prot{e} = (A/C)\prot{\substrate} + (B/C)\prot{\product}$, we arrive at a simple product elimination rule: products of the form se are eliminated by using the linear elimination of the enzyme, multiplying by a factor of $\prot{s}$, and then identifying $\prot{s\substrate}=\prot{s}\prot{\substrate}$ and $\prot{s\product}=\prot{s}\prot{\product}$. It is straightforward to check that the same rule applies to the elimination of observables of type be and be$'$.
 
The only remaining fast observables that we need to eliminate are the linear bulk enzyme concentrations. Their equations of motion from \eqref{eq:MMdimensionless} and \eqref{eq:MMfluxes_dimensionless} are
\begin{equation}
\begin{split}
   \frac{\partial}{\partial t}\prot{\enzyme} &= -(\fast/\y{\enzyme})\left[
      \km{c}{\substrate}{\enzyme}\y{\enzyme} \prot{\enzyme} +
      \kp{\substrate}{\enzyme}{c}\y{\substrate}\y{\enzyme}
      (\prot{\substrate}  + \prot{\enzyme} + \prot{\substrate\enzyme})\right.\\
&\quad+
\left.
      \km{c}{\product}{\enzyme}\y{\enzyme}\prot{\enzyme} +
      \kp{\product}{\enzyme}{c}\y{\product}\y{\enzyme}
      (\prot{\product} + \prot{\enzyme} + \prot{\product\enzyme})
\right]
 \end{split}
 \end{equation}
These contain product variables of the form se and be, which can now be eliminated using the method above, giving expressions in the form of equation \eqref{eq:se_elimination}. Substituting these and solving the quasi-steady state condition $(\partial/\partial t)\prot{\enzyme}=0$ then gives 
\begin{equation}
  \begin{split}
  \prot{\enzyme} &= -\frac{1}{(\km{c}{\substrate}{\enzyme} + \kp{\substrate}{\enzyme}{c}\y{\substrate} + \km{c}{\product}{\enzyme} + \kp{\product}{\enzyme}{c}\y{\product})^2}
\left[ \kp{\substrate}{\enzyme}{c}\y{\substrate}\Big((\km{c}{\substrate}{\enzyme} + \kp{\substrate}{\enzyme}{c}\y{\substrate}
 + \km{c}{\product}{\enzyme} + \kp{\product}{\enzyme}{c}\y{\product})\prot{\substrate}\right.\\
&\quad + \kp{\substrate}{\enzyme}{c}\y{\substrate}\prot{\substrate\substrate} + \kp{\product}{\enzyme}{c}\y{\product}\prot{\substrate\product}\Big)\\
&\quad +\left. \kp{\product}{\enzyme}{c}\y{\product}\Big((\km{c}{\substrate}{\enzyme} + \kp{\substrate}{\enzyme}{c}\y{\substrate} + \km{c}{\product}{\enzyme} + \kp{\product}{\enzyme}{c}\y{\product})\prot{\product}
+ \kp{\substrate}{\enzyme}{c}\y{\substrate}\prot{\substrate\product} + \kp{\product}{\enzyme}{c}\y{\product}\prot{\product\product}\Big)\right]    
   \end{split}
\end{equation}

We can now compare to the standard Michaelis-Menten elimination of the bulk enzyme, which treats products like $\prot{\substrate\enzyme}$ not as separate observables but identifies them with $\prot{\substrate}\prot{\enzyme}$ and then solves $(\partial/\partial t)\prot{\enzyme} = 0$. It is straightforward to check that our above elimination formula is just this Michaelis-Menten result expanded to quadratic order. We will therefore call this result ``quadratic quasi-steady state elimination''.

The result of the first stage of elimination is therefore that we can construct equations of motion for s, e$'$, b, sb and bb by quadratic quasi-steady state elimination of the bulk enzymes. The full quadratic elimination is not needed for all observables as the equations of motion for sb and bb only contain quadratic observables: in these we use linear enzyme elimination to replace products as explained above.

We can now write down explicitly what the effective contributions to the equations of motion for a substrate and product are. One starts from \eqref{eq:MMdimensionless} and \eqref{eq:MMfluxes_dimensionless} again,
\begin{equation}
   \frac{\partial}{\partial t}\prot{\substrate} = -(1/\y{\substrate})\left[
      \km{c}{\substrate}{\enzyme}\y{\enzyme} \prot{\enzyme}
+    \kp{\substrate}{\enzyme}{c}\y{\substrate}\y{\enzyme}
      (\prot{\substrate}  + \prot{\enzyme} + \prot{\substrate\enzyme})
\right] + \ldots
\end{equation}
and substitutes in the elimination formulae for $\prot{\enzyme}$ and $\prot{\substrate\enzyme}$. After a little algebra, the 
effective equation of motion for the substrate, and analogously the product, can be written in terms of unary reactions, as was the case for the linearised dynamics:
\begin{equation}
  \label{eq:MMnonlin_eff}
\begin{split}
      \frac{\partial}{\partial t}\prot{\substrate} &= -\hat{\lambda}_{\substrate\product}\prot{\substrate} + \hat{\lambda}_{\product\substrate}(\y{\product}/\y{\substrate})\prot{\product}
     + \ldots\\
    \frac{\partial}{\partial t}\prot{\product} &= -\hat{\lambda}_{\product\substrate}\prot{\product} + \hat{\lambda}_{\substrate\product}(\y{\substrate}/\y{\product})\prot{\substrate} + \ldots
\end{split}
\end{equation}
The difference is that the reaction rates $\hat{\lambda}_{\substrate\product}$ and $\hat{\lambda}_{\product\substrate}$ are now linearly dependent on substrate and product concentrations. In terms of the (constant) reaction rates $\lambda_{\product\substrate}$ and $\lambda_{\substrate\product}$ defined in \eqref{eq:unary_ratesMM}, we can write this concentration dependence in the simple form
\begin{equation}
  \label{eq:prod_ratesMM}
  \begin{split}
\hat{\lambda}_{\substrate\product} &= \lambda_{\substrate\product}\left(1-\frac{\prot{\substrate}\y{\substrate}/\Kms + \prot{\product}\y{\product}/\Kmp}{1 + \y{\substrate}/\Kms + \y{\product}/\Kmp}\right)\\
\hat{\lambda}_{\product\substrate} &= \lambda_{\product\substrate}\left(1-  \frac{\prot{\substrate}\y{\substrate}/\Kms + \prot{\product}\y{\product}/\Kmp}{1 + \y{\substrate}/\Kms + \y{\product}/\Kmp}\right)
  \end{split}
\end{equation}
Representing every Michaelis-Menten term in the bulk or on the boundary in this form, we thus obtain a set of equations from which all bulk enzymes have been eliminated. The coefficients in these equations then define the effective $L$-matrix $\Leff$ (or more precisely those columns of it that we use to obtain the rate and memory matrix).

\subsection{Elimination of fast subnetwork observables}

The result of the first stage of elimination is a rate matrix and memory matrix for the projected equations of the subnetwork variables s and e$'$. The second and final stage in the elimination of fast observables is now to remove the subnetwork enzymes e$'$.

This second stage is relatively simple because the projected equations of motion for e$'$ observables are in fact just the original mass action equations. This is so because the equations of motion for e$'$ observables only contain fast contributions from e$'$ and se$'$, so do not couple to any variables that were eliminated in the first stage. Furthermore, e$'$ observables are interior subnetwork species and do not couple to the slow bulk variables. Therefore their equations of motion cannot acquire memory terms. Eliminating subnetwork enzymes is then trivial: in the limit $\fast\to\infty$ the quasi-steady state assumption will become exact for them. Solving this gives exactly the standard Michaelis-Menten expression for the enzyme concentrations.

The final question is then where the subnetwork enzymes e$'$ feature in the projected equations for slow subnetwork species s. They appear in the rate matrix and substituting them there simply produces the usual Michaelis-Menten terms, in their full nonlinear form rather than expanded to second order as for bulk enzymes.

In the memory function e$'$ cannot appear linearly as subnetwork enzymes are not on the boundary. Such enzymes could then only appear in the memory via se$'$ terms. For this to occur would require nonzero entries in the blocks $\Leff^{\rm{se'},\rm{b}}$, $\Leff^{\rm{se'},\rm{sb}}$ or $\Leff^{\rm{se'},\rm{bb}}$, which are the relevant pieces of the leftmost factor $\lblock{S}{B}$ in \eqref{eq:memLT}. The block $\Leff^{\rm{se'},\rm{b}}$ contains the nonlinear se$'$ contributions in the equation of motion for a bulk observable; however such an equation cannot involve any se$'$ products because subnetwork enzymes do not interact with the bulk. Therefore there are no contributions to $\Leff^{\rm{se'},\rm{b}}$. Similarly $\Leff^{\rm{se'},\rm{bb}}$ must be zero as from the product rule of differentiation in \eqref{eq:product_diff} there would have to be a shared species in the first and second product index in order to obtain a nonzero contribution. For the block $\Leff^{\rm{se'},\rm{sb}}$ to have nonzero entries there must be a shared index, which in this case must be s. This means that nonzero elements could come only from the linearised equation of motion for the b observable; however, there are no contributions from e$'$ in the equations of motion for bulk observables and therefore $\Leff^{\rm{se'},\rm{sb}}$ must also be zero. In summary, this means that there are no contributions from subnetwork enzymes to any memory function.

\subsection{Summary of enzyme elimination procedure for nonlinear dynamics}
\label{subsec:summary_nonlin}

The final procedure we have arrived at for constructing projected equations for reaction systems with Michaelis-Menten terms, for nonlinear dynamics, can be split into two simple steps. The first is to construct the reduced $L$-matrix $\Leff$ by expanding all the Michaelis-Menten terms to second order around the steady state. From this matrix we are then able to calculate the rate matrix and memory function. Once this is done, we reinstate the full nonlinear form of the Michaelis-Menten terms from enzymes in the subnetwork.

We note as an aside that as an alternative to the above method, one could assign the subnetwork enzymes e$'$ to the bulk in the projection, on the grounds that one does not want to track them explicitly in the final projected equations. This would have the advantage of removing the need for a second stage of fast variable elimination. On the other hand, even though the subnetwork enzymes are fast variables, assigning them to the bulk and projecting remains an approximation because it would result in an additional contribution to the random force that we would discard in using the projected equations in practice. More intuitively, the approximation arises from the fact that in (\ref{eq:Lmatrixeqn}) we are neglecting cubic terms in $\delta x$. We therefore prefer to treat the subnetwork enzymes explicitly and to eliminate them separately in the second stage of the process. As this only requires re-instating the original Michaelis-Menten terms in the subnetwork, it is well worth doing in order to exclude one source of potential approximation error.

\section{Numerical Comparisons}
\label{sec:MMcomparison}

\begin{figure}[!ht]
  \centering
 \includegraphics[scale=0.7]{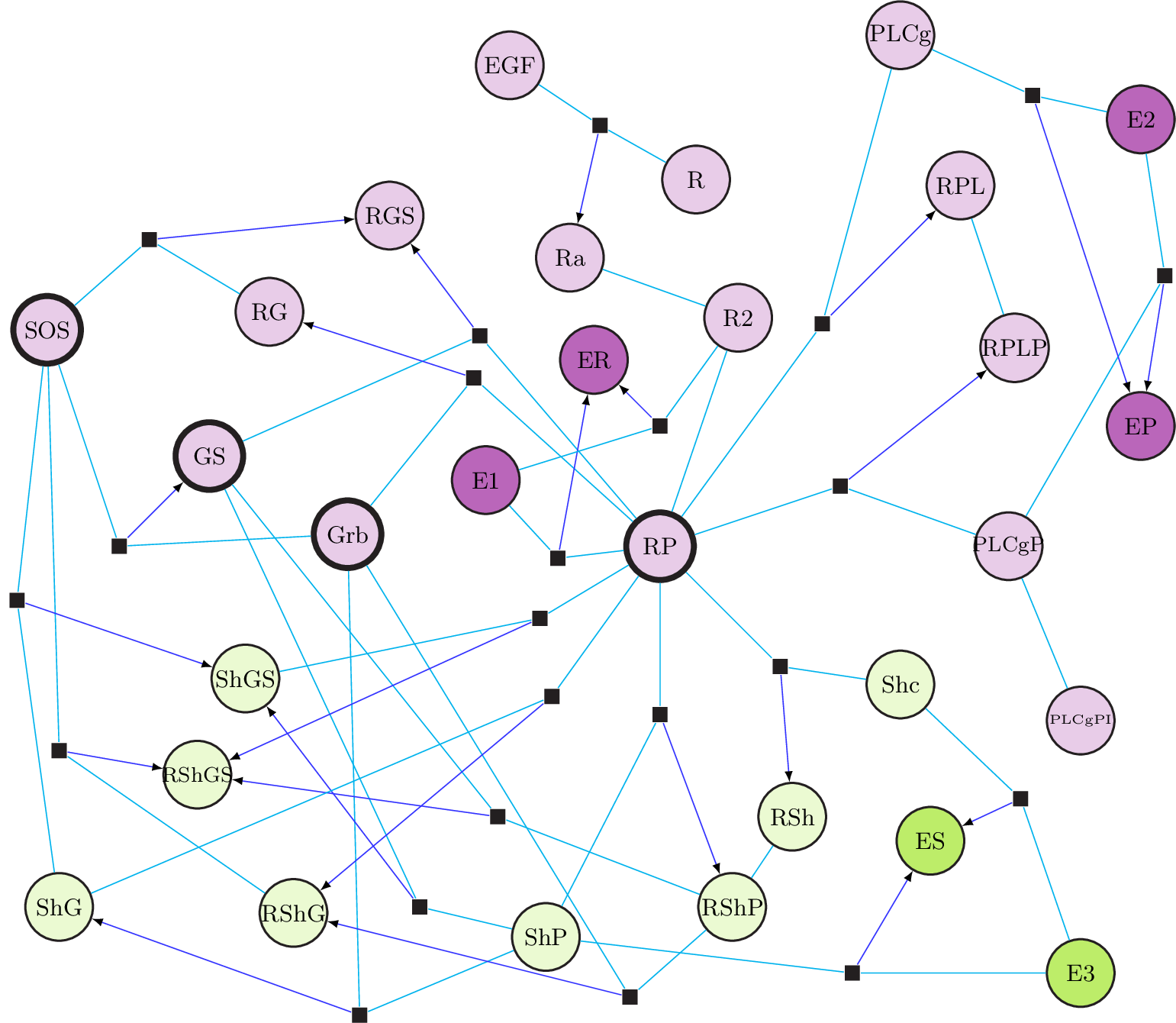}
  \caption[EGFR network with highlighted enzymes]{EGFR network as described in
    \citet{Kholodenko99}, adapted to include enzyme reactions. Three added enzyme reactions (highlighted by darker colours) with enzymes denoted E1-3 and enzyme-substrate complexes ``enzyme-R'' (denoted ER), ``enzyme-PLC$\gamma$'' (denoted EP), ``enzyme-Shc'' (denoted ES) capture the Michaelis-Menten contributions to the dynamics. Lines directly connecting molecular species indicate unary conversion reactions. We show binary reactions in the style of a factor graph representation \cite{Bishop2006}: each reaction is represented by a square; the two reaction partners are connected to the square by lines and an arrow points from the square to the reaction product. (The arrow does not have a meaning beyond this, and in particular does not indicate the direction in which the reaction takes place.) Nodes in the chosen subnetwork are coloured pink and purple; the four heavy circles indicate the boundary nodes.}
  \label{fig:MMegfr}
\end{figure}

We test our results on a model of the signalling network of epidermal growth factor receptor (EGFR) developed by \citet{Kholodenko99}. It is a well-studied network and contains a number of subnetworks. The EGFR network model as shown in Fig.~\ref{fig:MMegfr} has two subnetwork Michaelis-Menten reactions and one bulk Michaelis-Menten reaction. We choose the bulk to be the protein Shc and all complexes that include Shc.

In this section our aim is to compare two versions of the projected equations: the ones obtained by adding enzyme species explicitly with finite but large enzyme reaction rate constants, and the ones we get by eliminating enzymes in closed form as explained above. We will compare both projected descriptions with the 
dynamics of the full EGFR network \cite{Kholodenko99}, i.e.\ tracking explicitly the bulk degrees of freedom, to see which of the two better represents the true time courses. The bulk is assumed to be at steady state initially. For the subnetwork we choose initial conditions that maximize nonlinear effects but still give the desired steady state (i.e.\ have the appropriate value for all conserved quantities.)
We use $\delta=\left(\sum_{i=1}^N [\prot{i}(0)]^2/N\right)^{1/2}$ to quantify the initial deviation from the steady state, where $N$ is the number
of subnetwork species. 
To assess the accuracy of our two approximations we define
\begin{equation}
  \label{eq:errormeasure}
  \Delta=\frac{1}{T}\int_0^Tdt'\,\frac{1}{N}\sum_{i=1}^N|\prot{i}(t)-\delta\hat{x}_i(t)|
\end{equation}
where $\delta\hat{x}$ is our approximation to the subnetwork time course and  the total time interval for the error measurement is chosen such as to capture the transient dynamics in the approach to the overall system steady state, with $T=150$s.

To apply the projection method we analytically obtain all the relevant parts of $\bm{L}$ using the reaction rate constants and steady state values \cite{Rubin2014d}. Applying the method summarized in Sec.~\ref{subsec:summary_nonlin}, we then find the projected equations: a set of dynamical equations for the subnetwork species involving memory terms. These integro-differential equations can in principle be integrated directly, but as numerical algorithms for such equations are not in widespread use, it is advantageous to transform them to a system of differential equations for an enlarged set of variables. These can then be integrated using any standard differential equation solver.

The transformation is achieved by diagonalizing the matrix $\lblockb{B}{B}$ in the matrix exponential in \eqref{eq:mem}; we denote its eigenvalues, which should have negative real part, by $-\lambda_a$. Decomposing the matrix exponential into the contributions from these eigenvalues and corresponding (left and right) eigenvectors, and multiplying from the left and right by $
\lblockb{S}{B}$ and $\lblockb{B}{S}$ respectively, one can write the memory function in the form
\begin{equation}
M_{\beta\alpha}(\dt) = \sum_a
m_{\beta\alpha}^ae^{-\lambda_a \dt}
\end{equation}
where the $m_{\beta\alpha}^a$ are constant coefficients. The projected equations \eqref{eq:projectedeqns} can then be expressed as
\begin{equation}
  \label{eq:projectedeqnssolve}
  \frac{\partial}{\partial t}a_\alpha(t) = \sum_\beta a_\beta(t)\Omega_{\beta\alpha} + \sum_a c_\alpha^a(t) + r_\alpha(t)
\end{equation}
with
\begin{equation}
  c_\alpha^a(t) = \int_0^tdt'\,\sum_\beta a_\beta(t')m_{\beta\alpha}^ae^{-\lambda_a(t-t')}
\end{equation}
The memory terms that appear are now pure exponentials, so the $c_\alpha^a$ can instead be obtained by solving the {\em differential} equations
\begin{equation}
\frac{\partial}{\partial t} c_\alpha^a(t) = - \lambda_a c_\alpha^a(t) + \sum_\beta a_\beta(t)m_{\beta\alpha}^a
\label{eq:c_eqns}
\end{equation}
with initial condition $c_\alpha^a(0)=0$. In our numerics we thus solve (\ref{eq:projectedeqnssolve},\ref{eq:c_eqns}), with the random force term $r_\alpha(t)$ omitted as usual. The solution gives us the desired time courses of the subnetwork species.


%

\subsection{Explicit enzyme reactions}
\label{sec:EGFRMMtoMA}

For the projected equations with enzymes represented explicitly, we need to convert the three (irreversible) Michaelis-Menten reactions in the EGFR network model \cite{Kholodenko99},
\begin{equation}
  \begin{split}
    \text{RP} &\rightleftharpoons \text{R}_2\\
    \text{PLC}\gamma\text{P}&\rightleftharpoons \text{PLC}\gamma\\
    \text{ShP}&\rightleftharpoons \text{Shc}
  \end{split}
\end{equation}
into mass action form. The reversible harpoons above are intended to indicate only that there is a Michaelis-Menten reaction between the two species, and not e.g.\ whether it is reversible or irreversible. For each such reaction we use the relevant $\Km$ and $\vmax$ values together with a suitable value of the fast rate scale $\fast$ to create a set of mass action reaction rate constants. The relevant mass action terms are then added to the equations for substrate and product, and we add a mass action equation for the time evolution of the concentrations of enzyme and enzyme complex, as in \eqref{eq:MMmassaction}. The steady states of the enzyme and enzyme complex, which we require for the construction of the projected equations, can then be found by solving the relevant equations for the substrate, enzyme and enzyme complex. We note that to solve the full system of equations, the steady state values of the enzyme-substrate complexes must be added to the relevant conservation laws to ensure that the correct steady state is reached, with the same concentrations as in the Michaelis-Menten description. From this explicit expanded mass action system we can then construct projected equations, either linearised or nonlinear, in the standard manner explained in Sec.~\ref{sec:MMproj}.

\subsection{Consistency of enzyme elimination in linearised dynamics}

To eliminate the enzymes from the projected equations, we apply the elimination procedure as described for linearised dynamics in Section \ref{sec:linMM}: we write the enzyme reactions as unary reactions (effectively conformational changes) between substrate and product and construct from the resulting $\Leff$ a set of projected equations for the subnetwork observables that no longer makes explicit reference to enzymes.

\begin{figure}[!ht]
  \centering
\includegraphics[scale=0.8]{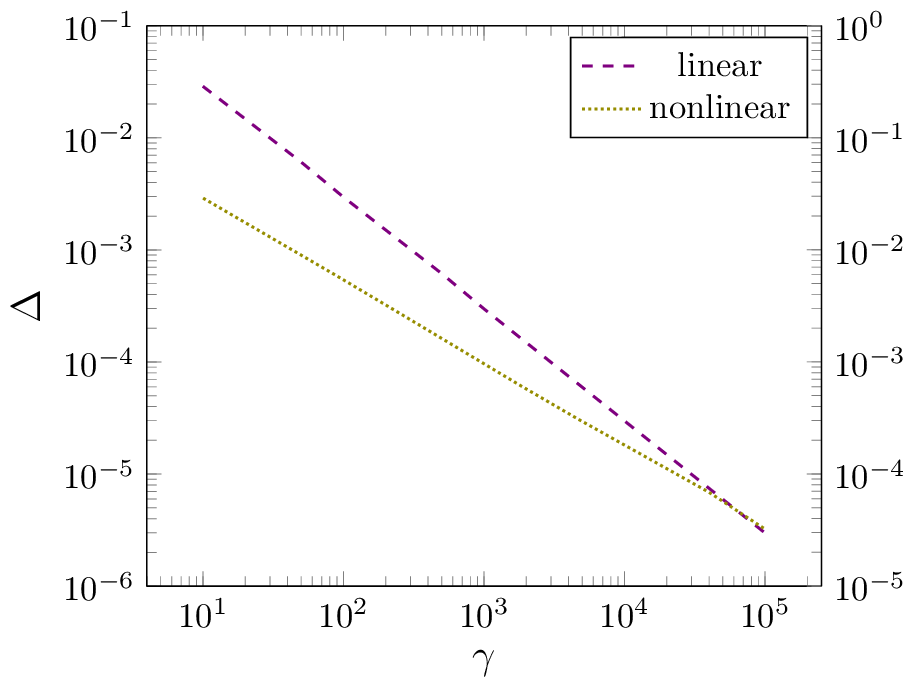}
  \caption[Errors from enzymes in linearised and nonlinear dynamics]{Deviations between predictions of projected equations with explicit enzymes and with enzymes eliminated, as a function of the fast enzyme rate scale $\fast$ used when enzymes are represented explicitly. Shown are results for an initial condition far from the steady state, with $\delta \approx 2$. The two curves show the deviations for the two separate cases of linearised and nonlinear dynamics. 
  }
  \label{fig:MMlin_nonlin_errors}
\end{figure}

Fig.~\ref{fig:MMlin_nonlin_errors} shows how the predictions from the linearised projected equations with enzymes represented explicitly approach the ones from enzyme elimination as the fast enzyme reaction scale $\fast$ grows. Specifically we show the deviation measure (\ref{eq:errormeasure}) that is obtained if we insert for $\delta x_i$ and $\delta \hat x_i$ the predictions from the two types of projected equations.  
As $\fast$ increases, the deviation decays to zero as it should. This provides an important consistency check on our enzyme elimination method, which we obtained by taking {\em analytically} the large $\fast$ limit of the projected equations with explicit enzymes.

\subsection{Consistency of enzyme elimination in nonlinear dynamics}

For the full nonlinear dynamics we derive the projected equations with enzymes eliminated using the method described in Sec.~\ref{sec:nonlinMM}. As for the linearised dynamics we use the predictions from these equations as a baseline in a comparison against the projected equations for the reaction network with explicit enzymes. Figure \ref{fig:MMlin_nonlin_errors} shows that the deviation between the time courses predicted by the two sets of equations again decreases towards zero with increasing $\fast$ as it should. 

\subsection{Accuracy of projected equations}

\begin{figure}[!ht]
  \centering
\includegraphics[scale=0.8]{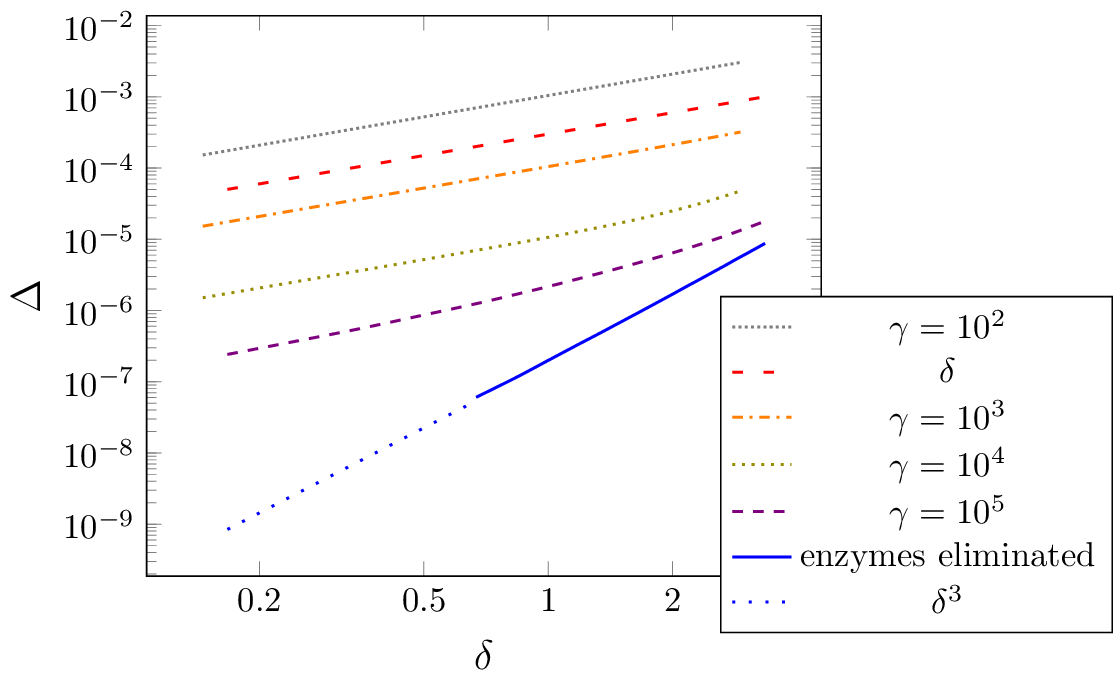}
  \caption[Errors from enzymes in nonlinear dynamics]{Log-log plot of approximation error $\Delta$ vs initial deviation from steady state $\delta$. We compare the nonlinear projected equations with enzymes eliminated and with enzymes represented explicitly for different values of the fast rate scale parameter $\fast$. The baseline in both cases are the time courses from the full system of reaction equations including Michaelis-Menten terms (i.e.\ with the quasi-steady state approximation, which is exact for $\gamma\to\infty$). The enzyme elimination method has an approximation error growing as $\delta^3$ as indicated by the blue dotted line; for the smallest $\delta$ the errors become too small to measure accurately.
The projected equations with explicit enzymes produce much larger errors that grow linearly in $\delta$ for small $\delta$ 
. 
}
  \label{fig:MMnonlineqnerrors}
\end{figure}

While the results in Fig.~\ref{fig:MMlin_nonlin_errors} are useful as consistency checks, the question that we are primarily interested in is how well the projected equations perform in predicting the real dynamics of the chosen subnetwork, as given by the full system of reaction equations including Michaelis-Menten terms. 
Fig.~\ref{fig:MMnonlineqnerrors} shows the approximation errors $\Delta$ of the predictions made by (a) the simpler projected equations, with enzymes explicitly dependent and therefore involving a fast enzyme rate scale parameter $\fast$, and (b) the projected equations derived using the new enzyme elimination method. In both cases we consider the full nonlinear projected equations, and show the approximation error $\Delta$ as a function of the distance $\delta$ of the initial conditions from the steady state.

One can argue on general grounds that nonlinear projected equations should have an approximation error growing only as $\delta^3$, whereas simpler approximation methods give much larger errors of order $\delta$~\cite{Rubin2014d}. Fig.~\ref{fig:MMnonlineqnerrors} shows that the expectation of a $\delta^3$-scaling is obeyed well for the projected equations derived using the enzyme elimination method. The projected equations derived by representing enzymes explicitly, on the other hand, have significantly higher $\Delta$, even for the largest $\fast=10^5$ that we can use while still being able to solve the resulting stiff differential equations. The approximation error scales $\Delta \sim \delta$ over a large range, indicating that the introduction of the explicit enzymes and their reaction rates causes a much larger approximation error than the projection method (with its neglect of the random force) itself. One would expect from the trends in Fig.~\ref{fig:MMnonlineqnerrors} that for larger $\fast$ than we can access, the error would grow as $\sim \delta$ for small $\delta$ but then cross over to the $\fast\to\infty$ (enzyme elimination) curve at larger $\delta$. The crossover value of $\delta$ has to go to zero as $\fast$ grows to reflect the consisteny between the $\fast\to\infty$ limit and the enzyme elimination method.


It is notable that computation times required to integrate the system of projected equations derived by enzyme elimination are shorter by roughly a factor of 2 than those for the case where enzymes are represented explicitly; this is due to the appearance of ``stiff'' terms in the latter case, which the enzyme elimination avoids. The enzyme elimination method thus is not only more accurate but also computationally more efficient.

\section{Discussion}

We have considered the problem of describing subnetwork dynamics in protein interaction networks. The projection approach we have previously developed gives accurate results in this regard, but can be applied directly only to systems with unary and binary reactions. Our aim in this paper was to extend it to systems involving enzymatic reactions represented as Michaelis-Menten terms.

We summarized the conditions under which enzyme reactions represented in mass action form become equivalent to Michaelis-Menten kinetics. This suggested  a convenient scaling, of fast enzyme reaction rates and simultaneously low enzyme concentrations, that exactly reproduces Michaelis-Menten equations in the limit where the relevant fast rate scale parameter $\fast$ grows large.

Applying this construction, one can map a protein interaction network with Michaelis-Menten reactions to an extended one with only unary and binary reactions; the main task is then to understand what limit is approached in the projected description when $\fast\to\infty$. By analysing where the fast enzyme degrees of freedom feature in the rate matrix and memory functions that define the projected equations, we showed that it is possible to construct the projected equations directly in the large $\fast$-limit, both for linearised and nonlinear dynamics, using a quasi-steady state elimination. This gives us effective unary reaction contributions to represent the Michaelis-Menten dynamics, with concentration-dependent reaction rate constants in the nonlinear case, and so allows one to constuct the projected equations without ever introducing enzymes explicitly.

The resulting method significantly widens the range of biochemical reaction systems to which the projection approach can be applied; as one example we showed an application to a subnetwork of the EGFR reaction network of \citet{Kholodenko99}. Here we demonstrated that the enzyme elimination method produces significantly more accurate predictions for the subnetwork dynamics than can be obtained when enzymes are represented explicitly, even for the largest values of the fast rate scale parameter $\fast$. We also found that enzyme elimination leads to gains in computational efficiency.

Our general approach to constructing projected equations for networks of nonlinear reaction equations should be more widely applicable still, and we hope in future work to consider e.g.\ Michaelis-Menten reactions with inhibition and Hill equations \cite{Murray2001}. Conceptually more demanding will be an extension to cases where copy number fluctuations from finite reaction volumes are significant, which would correspond to keeping nonzero $\epsilon$ in our notation. An elegant treatment of the case where the bulk contains only fast variables has been given in \citet{Thomas2010} and may serve as a useful point of departure in this direction.

\section*{Acknowledgements}

PS acknowledges the stimulating research environment provided by the EPSRC Centre for Doctoral Training in Cross-Disciplinary Approaches to Non-Equilibrium Systems (CANES, EP/L015854/1). KJR gratefully acknowledges a
BBSRC Quota Doctoral Training Grant.
\bibliography{mm_dynamics}

\begin{thebibliography}{20}%
\makeatletter
\providecommand \@ifxundefined [1]{%
 \@ifx{#1\undefined}
}%
\providecommand \@ifnum [1]{%
 \ifnum #1\expandafter \@firstoftwo
 \else \expandafter \@secondoftwo
 \fi
}%
\providecommand \@ifx [1]{%
 \ifx #1\expandafter \@firstoftwo
 \else \expandafter \@secondoftwo
 \fi
}%
\providecommand \natexlab [1]{#1}%
\providecommand \enquote  [1]{``#1''}%
\providecommand \bibnamefont  [1]{#1}%
\providecommand \bibfnamefont [1]{#1}%
\providecommand \citenamefont [1]{#1}%
\providecommand \href@noop [0]{\@secondoftwo}%
\providecommand \href [0]{\begingroup \@sanitize@url \@href}%
\providecommand \@href[1]{\@@startlink{#1}\@@href}%
\providecommand \@@href[1]{\endgroup#1\@@endlink}%
\providecommand \@sanitize@url [0]{\catcode `\\12\catcode `\$12\catcode
  `\&12\catcode `\#12\catcode `\^12\catcode `\_12\catcode `\%12\relax}%
\providecommand \@@startlink[1]{}%
\providecommand \@@endlink[0]{}%
\providecommand \url  [0]{\begingroup\@sanitize@url \@url }%
\providecommand \@url [1]{\endgroup\@href {#1}{\urlprefix }}%
\providecommand \urlprefix  [0]{URL }%
\providecommand \Eprint [0]{\href }%
\providecommand \doibase [0]{http://dx.doi.org/}%
\providecommand \selectlanguage [0]{\@gobble}%
\providecommand \bibinfo  [0]{\@secondoftwo}%
\providecommand \bibfield  [0]{\@secondoftwo}%
\providecommand \translation [1]{[#1]}%
\providecommand \BibitemOpen [0]{}%
\providecommand \bibitemStop [0]{}%
\providecommand \bibitemNoStop [0]{.\EOS\space}%
\providecommand \EOS [0]{\spacefactor3000\relax}%
\providecommand \BibitemShut  [1]{\csname bibitem#1\endcsname}%
\let\auto@bib@innerbib\@empty
\bibitem [{\citenamefont {Bhalla}(2003)}]{Bhalla2003}%
  \BibitemOpen
  \bibfield  {author} {\bibinfo {author} {\bibfnamefont {U.~S.}\ \bibnamefont
  {Bhalla}},\ }\href@noop {} {\bibfield  {journal} {\bibinfo  {journal} {Prog.
  Biophys. Mol. Bio.}\ }\textbf {\bibinfo {volume} {81}},\ \bibinfo {pages}
  {45} (\bibinfo {year} {2003})}\BibitemShut {NoStop}%
\bibitem [{\citenamefont {Ackermann}\ \emph {et~al.}(2012)\citenamefont
  {Ackermann}, \citenamefont {Einloft}, \citenamefont {N\"{o}then},\ and\
  \citenamefont {Koch}}]{Ackermann2012}%
  \BibitemOpen
  \bibfield  {author} {\bibinfo {author} {\bibfnamefont {J.}~\bibnamefont
  {Ackermann}}, \bibinfo {author} {\bibfnamefont {J.}~\bibnamefont {Einloft}},
  \bibinfo {author} {\bibfnamefont {J.}~\bibnamefont {N\"{o}then}}, \ and\
  \bibinfo {author} {\bibfnamefont {I.}~\bibnamefont {Koch}},\ }\href {\doibase
  http://dx.doi.org/10.1016/j.jtbi.2012.08.042} {\bibfield  {journal} {\bibinfo
   {journal} {J. Theor. Biol.}\ }\textbf {\bibinfo {volume} {315}},\ \bibinfo
  {pages} {71–80} (\bibinfo {year} {2012})}\BibitemShut {NoStop}%
\bibitem [{\citenamefont {Conradi}\ \emph {et~al.}(2007)\citenamefont
  {Conradi}, \citenamefont {Flockerzi}, \citenamefont {Raisch},\ and\
  \citenamefont {Stelling}}]{Conradi2007}%
  \BibitemOpen
  \bibfield  {author} {\bibinfo {author} {\bibfnamefont {C.}~\bibnamefont
  {Conradi}}, \bibinfo {author} {\bibfnamefont {D.}~\bibnamefont {Flockerzi}},
  \bibinfo {author} {\bibfnamefont {J.}~\bibnamefont {Raisch}}, \ and\ \bibinfo
  {author} {\bibfnamefont {J.}~\bibnamefont {Stelling}},\ }\href {\doibase
  10.1073/pnas.0705731104} {\bibfield  {journal} {\bibinfo  {journal} {Proc.
  Natl. Acad. Sci. U. S. A.}\ }\textbf {\bibinfo {volume} {104}},\ \bibinfo
  {pages} {19175} (\bibinfo {year} {2007})}\BibitemShut {NoStop}%
\bibitem [{\citenamefont {Okino}\ and\ \citenamefont
  {Mavrovouniotis}(1998)}]{Okino1998}%
  \BibitemOpen
  \bibfield  {author} {\bibinfo {author} {\bibfnamefont {M.~S.}\ \bibnamefont
  {Okino}}\ and\ \bibinfo {author} {\bibfnamefont {M.~L.}\ \bibnamefont
  {Mavrovouniotis}},\ }\href {\doibase 10.1021/cr950223l} {\bibfield  {journal}
  {\bibinfo  {journal} {Chem. Rev.}\ }\textbf {\bibinfo {volume} {98}},\
  \bibinfo {pages} {391} (\bibinfo {year} {1998})}\BibitemShut {NoStop}%
\bibitem [{\citenamefont {Radulescu}\ \emph {et~al.}(2012)\citenamefont
  {Radulescu}, \citenamefont {Gorban}, \citenamefont {Zinovyev},\ and\
  \citenamefont {Noel}}]{Radulescu2012}%
  \BibitemOpen
  \bibfield  {author} {\bibinfo {author} {\bibfnamefont {O.}~\bibnamefont
  {Radulescu}}, \bibinfo {author} {\bibfnamefont {A.~N.}\ \bibnamefont
  {Gorban}}, \bibinfo {author} {\bibfnamefont {A.}~\bibnamefont {Zinovyev}}, \
  and\ \bibinfo {author} {\bibfnamefont {V.}~\bibnamefont {Noel}},\ }\href
  {\doibase 10.3389/fgene.2012.00131} {\bibfield  {journal} {\bibinfo
  {journal} {Front. Genet.}\ }\textbf {\bibinfo {volume} {3}},\ \bibinfo
  {pages} {131} (\bibinfo {year} {2012})}\BibitemShut {NoStop}%
\bibitem [{\citenamefont {Rubin}\ \emph {et~al.}(2014)\citenamefont {Rubin},
  \citenamefont {Lawler}, \citenamefont {Sollich},\ and\ \citenamefont
  {Ng}}]{Rubin2014d}%
  \BibitemOpen
  \bibfield  {author} {\bibinfo {author} {\bibfnamefont {K.~J.}\ \bibnamefont
  {Rubin}}, \bibinfo {author} {\bibfnamefont {K.}~\bibnamefont {Lawler}},
  \bibinfo {author} {\bibfnamefont {P.}~\bibnamefont {Sollich}}, \ and\
  \bibinfo {author} {\bibfnamefont {T.}~\bibnamefont {Ng}},\ }\href {\doibase
  10.1016/j.jtbi.2014.06.002} {\bibfield  {journal} {\bibinfo  {journal} {J.
  Theor. Biol.}\ }\textbf {\bibinfo {volume} {357}},\ \bibinfo {pages} {245}
  (\bibinfo {year} {2014})}\BibitemShut {NoStop}%
\bibitem [{\citenamefont {Henri}(1902)}]{Henri1902}%
  \BibitemOpen
  \bibfield  {author} {\bibinfo {author} {\bibfnamefont {V.}~\bibnamefont
  {Henri}},\ }\href@noop {} {\bibfield  {journal} {\bibinfo  {journal} {C. R.
  Acad. Sci. Paris}\ }\textbf {\bibinfo {volume} {135}},\ \bibinfo {pages}
  {916} (\bibinfo {year} {1902})}\BibitemShut {NoStop}%
\bibitem [{\citenamefont {Michaelis}\ and\ \citenamefont
  {Menten}(1913)}]{Michaelis1913}%
  \BibitemOpen
  \bibfield  {author} {\bibinfo {author} {\bibfnamefont {L.}~\bibnamefont
  {Michaelis}}\ and\ \bibinfo {author} {\bibfnamefont {M.~L.}\ \bibnamefont
  {Menten}},\ }\href {\doibase http://web.lemoyne.edu/~giunta/menten.html}
  {\bibfield  {journal} {\bibinfo  {journal} {Biochemische Zeitschrift}\
  }\textbf {\bibinfo {volume} {49}},\ \bibinfo {pages} {333} (\bibinfo {year}
  {1913})}\BibitemShut {NoStop}%
\bibitem [{\citenamefont {Briggs}\ and\ \citenamefont
  {Haldane}(1925)}]{Briggs1925}%
  \BibitemOpen
  \bibfield  {author} {\bibinfo {author} {\bibfnamefont {G.~E.}\ \bibnamefont
  {Briggs}}\ and\ \bibinfo {author} {\bibfnamefont {J.~B.~S.}\ \bibnamefont
  {Haldane}},\ }\href@noop {} {\bibfield  {journal} {\bibinfo  {journal}
  {Biochem. J.}\ }\textbf {\bibinfo {volume} {19}},\ \bibinfo {pages} {338}
  (\bibinfo {year} {1925})}\BibitemShut {NoStop}%
\bibitem [{\citenamefont {Haldane}(1930)}]{Haldane1930}%
  \BibitemOpen
  \bibfield  {author} {\bibinfo {author} {\bibfnamefont {J.~B.~S.}\
  \bibnamefont {Haldane}},\ }\href@noop {} {\emph {\bibinfo {title}
  {Enzymes}}}\ (\bibinfo  {publisher} {Longmans, Green and Co., London},\
  \bibinfo {year} {1930})\BibitemShut {NoStop}%
\bibitem [{\citenamefont {Sauro}(2013)}]{Sauro2013}%
  \BibitemOpen
  \bibfield  {author} {\bibinfo {author} {\bibfnamefont {H.~M.}\ \bibnamefont
  {Sauro}},\ }\href@noop {} {\emph {\bibinfo {title} {Enzyme Kinetics for
  Systems Biology}}}\ (\bibinfo  {publisher} {Ambrosius Publishing},\ \bibinfo
  {year} {2013})\BibitemShut {NoStop}%
\bibitem [{\citenamefont {Segel}\ and\ \citenamefont
  {Slemrod}(1989)}]{Segel1989}%
  \BibitemOpen
  \bibfield  {author} {\bibinfo {author} {\bibfnamefont {L.}~\bibnamefont
  {Segel}}\ and\ \bibinfo {author} {\bibfnamefont {M.}~\bibnamefont
  {Slemrod}},\ }\href@noop {} {\bibfield  {journal} {\bibinfo  {journal} {SIAM
  Review}\ }\textbf {\bibinfo {volume} {31}},\ \bibinfo {pages} {446} (\bibinfo
  {year} {1989})}\BibitemShut {NoStop}%
\bibitem [{\citenamefont {Van~Slyke}\ and\ \citenamefont
  {Cullen}(1914)}]{VanSlyke1914}%
  \BibitemOpen
  \bibfield  {author} {\bibinfo {author} {\bibfnamefont {D.~D.}\ \bibnamefont
  {Van~Slyke}}\ and\ \bibinfo {author} {\bibfnamefont {G.~E.}\ \bibnamefont
  {Cullen}},\ }\href@noop {} {\bibfield  {journal} {\bibinfo  {journal} {J.
  Biol. Chem.}\ }\textbf {\bibinfo {volume} {19}},\ \bibinfo {pages} {141}
  (\bibinfo {year} {1914})}\BibitemShut {NoStop}%
\bibitem [{\citenamefont {Li}\ and\ \citenamefont {Li}(2013)}]{Li2013}%
  \BibitemOpen
  \bibfield  {author} {\bibinfo {author} {\bibfnamefont {B.}~\bibnamefont
  {Li}}\ and\ \bibinfo {author} {\bibfnamefont {B.}~\bibnamefont {Li}},\ }\href
  {\doibase 10.1007/s10910-013-0229-5} {\bibfield  {journal} {\bibinfo
  {journal} {J. Math. Chem.}\ }\textbf {\bibinfo {volume} {51}},\ \bibinfo
  {pages} {2668} (\bibinfo {year} {2013})}\BibitemShut {NoStop}%
\bibitem [{\citenamefont {Kollar}\ and\ \citenamefont
  {Siskova}(2015)}]{Kollar2015}%
  \BibitemOpen
  \bibfield  {author} {\bibinfo {author} {\bibfnamefont {R.}~\bibnamefont
  {Kollar}}\ and\ \bibinfo {author} {\bibfnamefont {K.}~\bibnamefont
  {Siskova}},\ }\href {\doibase 10.1007/s11538-015-0090-8} {\bibfield
  {journal} {\bibinfo  {journal} {Bull. Math. Bio.}\ }\textbf {\bibinfo
  {volume} {77}},\ \bibinfo {pages} {1401} (\bibinfo {year}
  {2015})}\BibitemShut {NoStop}%
\bibitem [{\citenamefont {Mori}(1965)}]{Mori1965}%
  \BibitemOpen
  \bibfield  {author} {\bibinfo {author} {\bibfnamefont {H.}~\bibnamefont
  {Mori}},\ }\href {\doibase 10.1143/PTP.33.423} {\bibfield  {journal}
  {\bibinfo  {journal} {Prog. Theor. Phys.}\ }\textbf {\bibinfo {volume}
  {33}},\ \bibinfo {pages} {423} (\bibinfo {year} {1965})}\BibitemShut
  {NoStop}%
\bibitem [{\citenamefont {Kholodenko}\ \emph {et~al.}(1999)\citenamefont
  {Kholodenko}, \citenamefont {Demin}, \citenamefont {Moehren},\ and\
  \citenamefont {Hoek}}]{Kholodenko99}%
  \BibitemOpen
  \bibfield  {author} {\bibinfo {author} {\bibfnamefont {B.~N.}\ \bibnamefont
  {Kholodenko}}, \bibinfo {author} {\bibfnamefont {O.~V.}\ \bibnamefont
  {Demin}}, \bibinfo {author} {\bibfnamefont {G.}~\bibnamefont {Moehren}}, \
  and\ \bibinfo {author} {\bibfnamefont {J.~B.}\ \bibnamefont {Hoek}},\
  }\href@noop {} {\bibfield  {journal} {\bibinfo  {journal} {J. Biol. Chem.}\
  }\textbf {\bibinfo {volume} {274}},\ \bibinfo {pages} {30169} (\bibinfo
  {year} {1999})}\BibitemShut {NoStop}%
\bibitem [{\citenamefont {Bishop}(2006)}]{Bishop2006}%
  \BibitemOpen
  \bibfield  {author} {\bibinfo {author} {\bibfnamefont {C.~M.}\ \bibnamefont
  {Bishop}},\ }\href@noop {} {\emph {\bibinfo {title} {Pattern Recognition and
  Machine Learning}}}\ (\bibinfo  {publisher} {Springer},\ \bibinfo {year}
  {2006})\BibitemShut {NoStop}%
\bibitem [{\citenamefont {Murray}(2001)}]{Murray2001}%
  \BibitemOpen
  \bibfield  {author} {\bibinfo {author} {\bibfnamefont {J.}~\bibnamefont
  {Murray}},\ }\href@noop {} {\emph {\bibinfo {title} {Mathematical Biology I.
  An Introduction}}}\ (\bibinfo  {publisher} {Springer, New York},\ \bibinfo
  {year} {2001})\BibitemShut {NoStop}%
\bibitem [{\citenamefont {Thomas}\ \emph {et~al.}(2010)\citenamefont {Thomas},
  \citenamefont {Straube},\ and\ \citenamefont {Grima}}]{Thomas2010}%
  \BibitemOpen
  \bibfield  {author} {\bibinfo {author} {\bibfnamefont {P.}~\bibnamefont
  {Thomas}}, \bibinfo {author} {\bibfnamefont {A.~V.}\ \bibnamefont {Straube}},
  \ and\ \bibinfo {author} {\bibfnamefont {R.}~\bibnamefont {Grima}},\
  }\href@noop {} {\bibfield  {journal} {\bibinfo  {journal} {J. Chem. Phys.}\
  }\textbf {\bibinfo {volume} {133}},\ \bibinfo {pages} {195101} (\bibinfo
  {year} {2010})}\BibitemShut {NoStop}%
\end{thebibliography}%

\end{document}